\documentclass[useAMS,usenatbib]{mn2e}
\usepackage{epsf}
\usepackage{amssymb}
\usepackage{graphicx}
\usepackage[usenames]{color}
\usepackage{fmtcount}

\newcommand{\ergss}{~ergs~s$^{-1}$}

\newcommand {\bc}{\begin {center}}
\newcommand {\ec}{\end {center}}\newcommand {\be}{\begin {equation}}
\newcommand {\ee}{\end {equation}}
\newcommand {\disp}{\displaystyle}

\title[The nature of AGN-driven perturbation in Perseus]{The nature and energetics of AGN-driven perturbations in the hot gas in the Perseus Cluster}
\author[Zhuravleva et
  al.]{I. Zhuravleva$^{1,2}$\thanks{zhur@stanford.edu},
  E. Churazov$^{3,4}$, P. Ar{\'e}valo$^{5}$,
  A. A. Schekochihin$^{6,7}$, W. R. Forman$^{8}$, \newauthor S. W. Allen$^{1,2,9}$, A. Simionescu$^{10}$, R. Sunyaev$^{3,4}$,  A. Vikhlinin$^{8}$, N. Werner$^{1,2}$  \\
$^1$Kavli Institute for Particle Astrophysics and Cosmology, Stanford University, 452 Lomita Mall, Stanford, California 94305-4085, USA\\
$^2$Department of Physics, Stanford University, 382 Via Pueblo Mall,
Stanford, California 94305-4060, USA\\
$^3$Max Planck Institute for Astrophysics, Karl-Schwarzschild-Strasse 1, D-85741 Garching, Germany\\
$^4$Space Research Institute (IKI), Profsoyuznaya 84/32, Moscow
117997, Russia\\
$^5$Instituto de F{\'i}sica y Astronom{\'i}a, Facultad de Ciencias, Universidad de Valpara{\'i}so, Gran Bretana N 1111, Playa Ancha,
Valpara{\'i}so, Chile\\
$^6$Rudolf Peierls Centre for Theoretical Physics, University of Oxford, 1 Keble Rd, Oxford OX1 3NP, UK\\
$^7$Merton College, University of Oxford, Merton St, Oxford OX1 4JD,
UK\\
$^{8}$Harvard-Smithsonian Center for Astrophysics, 60 Garden Street, Cambridge, Massachusetts 02138, USA\\
$^9$SLAC National
Accelerator Laboratory, 2575 Sand Hill Road, Menlo Park, California 94025, USA\\
$^{10}$Japan Aerospace Exploration Agency, 3-1-1 Yoshinodai, Sagamihara, Kanagawa 252-5210, Japan\\
}

\begin{document}

\date{Accepted .... Received ...}

\pagerange{\pageref{firstpage}--\pageref{lastpage}} \pubyear{2016}

\maketitle

\label{firstpage}

\begin{abstract}
Cores of relaxed galaxy clusters are often disturbed by AGN. Their Chandra observations revealed a wealth of structures induced by shocks, subsonic gas motions, bubbles of relativistic plasma, etc. In this paper, we determine the nature and energy content of gas fluctuations in the Perseus core by probing statistical properties of emissivity fluctuations imprinted in the soft- and hard-band X-ray images. About $80$ per cent of the total variance of perturbations on $\sim 8 - 70$ kpc scales in the inner region have an isobaric nature, i.e., are consistent with slow displacements of the gas in pressure equilibrium with ambient medium. Observed variance translates to the ratio of non-thermal to thermal energy of $\sim 13$ per cent. In the region dominated by weak ``ripples'', about half of the total variance is also associated with isobaric perturbations on scales ~ a few tens of kpc. If these isobaric perturbations are induced by buoyantly rising bubbles, then these results suggest that most of the AGN-injected energy should first go into bubbles rather than into shocks. Using simulations of a shock propagating through the Perseus atmosphere, we found that models reproducing the observed features of a central shock have more than $50$ per cent of the AGN-injected energy associated with the bubble enthalpy and only about $20$ per cent is carried away with the shock. Such energy partition is consistent with the AGN-feedback model, mediated by bubbles of relativistic plasma, and supports the importance of turbulence in the balance between gas heating and radiative cooling.
\end{abstract}
\begin{keywords}methods: observational-methods: statistical-techniques: image processing-galaxies: clusters: intracluster medium-X-rays: galaxies: clusters
\end{keywords}

\section{Introduction}

Cores of clusters of galaxies are often perturbed by powerful jets
from central supermassive black holes. Interacting with intracluster
gas, these jets inflate bubbles of relativistic plasma, which expel
the hot gas, producing cavities in the X-ray images of clusters
\citep[e.g.,][]{Boe93,Chu00,McN00,McN07,Fab12,Bir12,Hla12}. The initial rapid
expansion of the bubbles may drive shocks that propagate
through intracluster medium (ICM) and heat the gas
\citep[e.g.,][]{Fab03,For07,Ran15,Rey15}. With time, rapid expansion
decelerates and the bubbles continue to grow subsonically until they
start rising buoyantly in the cluster atmosphere, uplifting cool gas,
exciting gravity waves, and driving turbulence
\citep{Chu02,Omm04,Zhu14b}. One of the alternative scenarios is that the bubbles themselves are composed of hot thermal plasma and efficiently mix with the ICM heating the gas  \citep{Hil14,Sok15}.

The X-ray-brightest nearby galaxy cluster, the Perseus Cluster,
provides a textbook example of radio-mode AGN feedback. Deep {\it
  Chandra} observations of the cluster unveiled distinct perturbations
in the hot gas in the innermost $\sim 100$ kpc region, where the
physics of the gas is governed by the powerful central AGN
\citep[][]{Fab11}. Namely, we clearly see bubbles of relativistic
plasma, shocks, cold filaments and fluctuations induced by motions of
the gas. The most recent summary of the observed features in the
Perseus core can be found in \citet{Fab11}. The statistical analysis
of the fluctuations was presented in \citet{Zhu15}. In this paper, we attempt to understand how the outflow energy from the central AGN is partitioned between different observed phenomena.

Here we will describe perturbations in the hot gas through their
``effective'' equation of state (EoS), which measures correlations
between observed fluctuations of density $\delta n/n$ and temperature $\delta T/T$,
expressed as 
\be
\disp\frac{\delta T}{T}=(\zeta_i-1)\left(\disp\frac{\delta
  n}{n}\right),
\ee
where $\zeta_i$ is the effective adiabatic index. This
correlation characterizes fluctuations of $T$ and $n$ relative to their mean values at a given distance from the cluster center. It may not reflect the true EoS of the ICM. Following the simplest approach,
we will consider only three types of perturbations. If conduction is
suppressed, then slowly displaced gas (e.g., via gravity waves, subsonic
turbulence) retains its initial entropy and stays in pressure
equilibrium with the ambient gas. Therefore, with respect to this ambient
gas, such perturbations will appear isobaric
($\zeta_{\rm isob.}=0$). Any local
changes in gas entropy will also appear as isobaric
perturbations. Weak shocks (sound waves) with Mach number $M-1\ll 1$
do not change gas entropy and, therefore, result in perturbations that appear 
adiabatic with $\zeta_{\rm adiab.}=5/3$. There are also cavities in the diffuse gas
associated with bubbles of relativistic plasma, which, observationally, can be interpreted as variations of the thermal
  gas density at constant temperature, i.e., isothermal fluctuations
  with $\zeta_{\rm isoth.}=1$.

The X-ray emissivity per unit volume is
\be
f(x,y,z)=n^2\Lambda(T),
\ee
where $\Lambda(T)$ is the X-ray emissivity in a given energy band. In the
soft band (e.g., $0.5-4$ keV for a gas with temperature $2-10$ keV), X-ray
emissivity $\Lambda(T)$ does not vary much with the gas
temperature. In contrast, in a hard band (e.g., $4-8$ keV), the
emissivity is temperature-dependent
\citep[e.g.,][]{For07,Zhu15}. Therefore, each type of perturbation
mentioned above will appear differently in the soft- and hard-band
X-ray images. Adiabatic perturbations will have a larger amplitude in
the hard band than in the soft band, while the amplitude of isobaric
fluctuations will be large in the soft band and small, or even zero, in
the hard band, depending on the choice of energy bands and the
temperature of the cluster gas. Bubbles produce similar depressions
in the X-ray images in both bands.

In this paper, we apply a statistical approach to compare the
amplitude of emissivity fluctuations in two different energy bands in
the Perseus Cluster by measuring the power spectra of fluctuations in
both bands and their cross-spectrum. We aim to establish the nature of
the observed perturbations in the cluster core, which are induced by
the central AGN, and to constrain the energy content associated
with each type of perturbation. This work is accompaniad by a recent
analysis of fluctuations in the Virgo/M87 Cluster \citep{Are15} and is
the third paper in a series devoted to the statistical
analysis of fluctuations in Perseus \citep[see][]{Zhu14b,Zhu15}.

\section{Isobaric, adiabatic and isothermal fluctuations}
\label{sec:ampl}

\begin{figure}
\includegraphics[trim=0 170 0 90,width=0.49\textwidth]{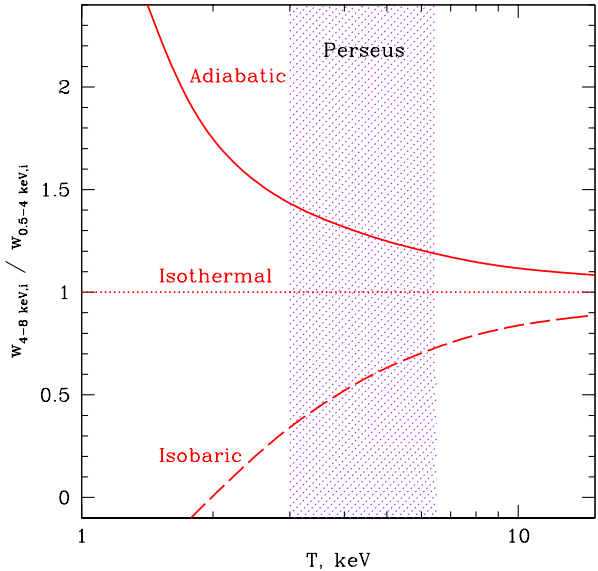}
\caption{The ratio (\ref{eq:flux_rat}) of emissivity perturbations in two energy bands, $0.5-4$ keV and $4-8$ keV, as a function of gas temperature, assuming that the perturbations are pure isobaric, isothermal or adiabatic. The purple dotted region shows the range of gas temperatures in
  the Perseus Cluster. Redshift, galactic HI column density and the abundance of heavy elements $Z=0.5Z_{\odot}$ (relative to the solar abundance of heavy elements, see \citet{And89}) of Perseus are used.
\label{fig:flux_nonlin}
}
\end{figure}

We denote by $\delta f/f$ the  emissivity fluctuation field in 3D, relative to the spatially smooth model.  Let us assume that  $\delta f/f$ can be
decomposed into isobaric, adiabatic and isothermal components,
\be
\frac{\delta f}{f}=\sum\limits_i\left(\frac{\delta f}{f}\right)_i,
\label{eq:sum}
\ee
where $i$ corresponds to one of the three considered types of perturbations. 
Of course, this trichotomy does not cover all
possibilities. For instance,  gas metallicity variations are not
captured by it. However, equation (\ref{eq:sum}) accounts for several major types of perturbations, which
are expected to be present in the ICM (see Discussion). If the typical
amplitude of fluctuations is small, one can link the emissivity
fluctuations to the corresponding density fluctuations. Namely, for the
$i$-th type of perturbations
\be
\left(\frac{\delta f}{f}\right)_i=\left(\frac{\delta n}{n}\right)_i\left[2+(\zeta_i-1)\frac{d\ln \Lambda(T)}{d\ln T}\right]\equiv\left(\frac{\delta n}{n}\right)_i w_i,
\ee
where $\zeta_i=0$, $5/3$ or $1$ for isobaric, adiabatic and isothermal fluctuations, respectively, and
\be
w_i=\left[2+(\zeta_i-1)\frac{d\ln \Lambda(T)}{d\ln T}\right].
\ee
For a given type of perturbations, the ratio of emissivity fluctuation fields measured in two different energy bands $a$ and $b$ is independent of density perturbations $(\delta n/n)_i$, namely,
\be
\disp\frac{(\delta f_b/f_b)_i}{(\delta f_a/f_a)_i}=\frac{w_{b,i}}{w_{a,i}}=\disp\frac{2+(\zeta_i-1)\disp\frac{d\ln
    \Lambda_b(T)}{d\ln T}}{2+(\zeta_i-1)\disp\frac{d\ln \Lambda_a(T)}{d\ln T}}.
\label{eq:flux_rat}
\ee
 Fig. \ref{fig:flux_nonlin} shows the ratio of emissivities in the
 soft ($0.5-4$ keV) and hard ($4-8$ keV) bands (the choice of the bands is justified in Appendix \ref{enbands}), assuming the abundance of heavy elements,
redshift and galactic HI column density of the Perseus Cluster. In a gas with the range of temperatures $3-6.5$
keV, characteristic for the Perseus core, $(\delta f_b/f_b)_i/(\delta f_a/f_a)_i\approx 1.3$, $1$
and $\approx 0.5$ for pure adiabatic, isothermal and isobaric
perturbations, respectively. Substantial difference between the curves
allows one to distinguish between different types of perturbations
using observations, which is more difficult to do for hotter
objects. 

\begin{figure*}
\includegraphics[trim=0 50 0 50,angle=0,width=\textwidth]{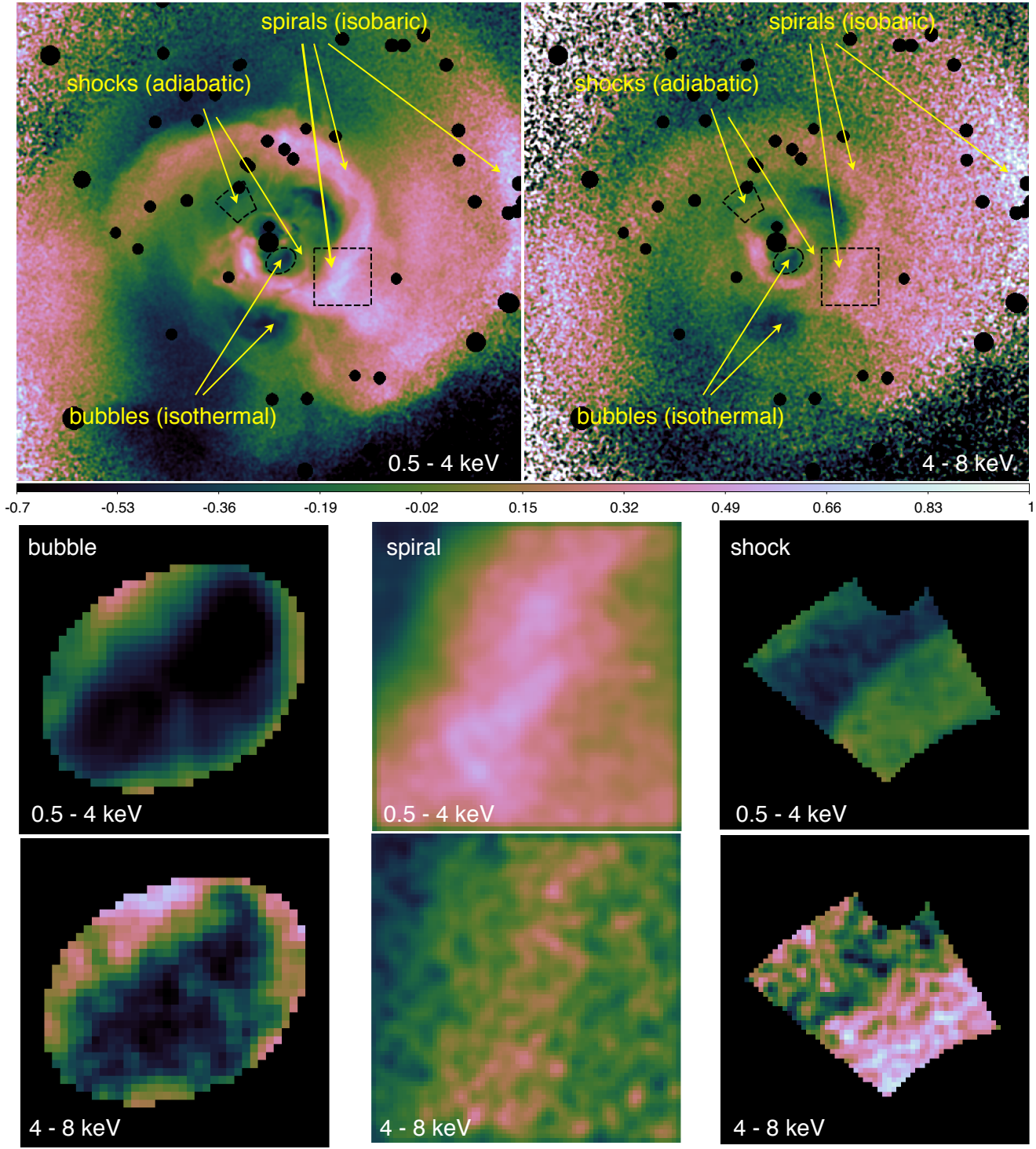}
\caption{{\it Chandra} images of the core ($\approx 200\times 200$ kpc, or
  $9.6\times9.6$ arcmin) of the Perseus Cluster divided by the
  spherically symmetric $\beta-$ model profiles in the 0.5 - 4 keV (top left) and
  4 - 8 keV (top right) bands. Point sources and the central AGN are
  excised. Arrows point to the prominent features identified in
  previous studies. Dashed regions indicate selected regions with a
  shock, a bubble and a spiral. They are enlarged in the middle and bottom
  panels. The color scales of the pair images are the same. For display
  purposes, all images are lightly smoothed with a $2''$ Gaussian.
\label{fig:images}
}
\end{figure*}

In practice, we are dealing not with the 3D emissivity fluctuation
field, but with the projection of this field onto the plane of the sky. However, for a nearly isothermal cluster, the ratio of surface
brightness fluctuations in two energy bands is expected to follow
equation (\ref{eq:flux_rat}), provided that a single type of
perturbations is dominating a region of interest (e.g., Churazov et
al., 2016, in prep.). The residual images (the initial images divided by the best-fitting
spherically symmetric $\beta$ models of the surface brightness) 
of the Perseus Cluster in $0.5-4$ keV and $4-8$ keV bands, Fig. \ref{fig:images} (see Section
\ref{sec:processing} for details), show that the amplitude
of the spiral-like feature is larger in the soft band than in the hard
band, hinting at an isobaric nature of this feature (see
Fig. \ref{fig:flux_nonlin}). However, in some regions around
the central bubbles, the amplitude of perturbations is larger in the
hard band, suggesting they have an adiabatic nature. Such visual examination allows us
to guess the nature of large-scale and large-amplitude
fluctuations. For the less prominent fluctuations that we cannot easily identify in the images, a different, statistical approach is needed.

 Assuming that $\delta f/f$ is a homogeneous and isotropic random
field, we calculate the cross spectrum (real part) $P_{k,ab}$ of emissivity
fluctuations in two energy bands $a$ and $b$ as
\be
P_{k,ab}=Re\left[\hat{\left(\frac{\delta f_a}{f_a}\right)}\cdot \hat{\left(\frac{\delta f_b}{f_b}\right)^*}\right],
\label{eq:psf}
\ee
where $k$ is a wavenumber and $ |\hat{(\delta f/f)} |$ is the Fourier transform of $\delta f/f$ in both bands. In practice, in order to avoid the adverse effects of non-periodic data and gaps in the exposure map, the calculations are done in real space using a modified $\Delta$ variance method to measure the surface brightness fluctuations in images in both bands \citep{Oss08,Are12}. First, fluctuations at a given spatial scale $1/k$ are singled out and then the variance of the convolved image is calculated, namely,
\be
  P_{k,ab}=\left\langle\left(\frac{\delta f_a}{f_a}\right)_{1/k} \left(\frac{\delta f_b}{f_b}\right)_{1/k}\right\rangle
\label{eq:ps}
\ee
where $\langle\rangle$ denotes
averaging over space, and $(\delta f_a/f_a)_{1/k}$ is the emissivity fluctuation field in the energy band $a$, filtered to keep only spatial scales $\sim 1/k$. Equation (\ref{eq:ps}) essentially provides an estimate of the conventional cross power spectrum [equation (\ref{eq:psf})] convolved with a smoothing kernel \citep[see, e.g.,][for details]{Are12,Chu12,Zhu14}.
Measuring power spectra $P_{k,aa}$ and  $P_{k,bb}$ in both bands and the cross spectrum $P_{k,ab}$, a scale-dependent correlation coefficient (coherence),
\be 
C(k)=\frac{P_{k,ab}}{\sqrt{P_{k,aa}P_{k,bb}}},
 \label{eq:coh}
\ee 
and a relative
amplitude (ratio) of fluctuations in two energy bands\footnote{As the
  number of photons in the soft band is often larger than in the hard band,
  the power spectrum of fluctuations in the soft band $P_{k,aa}$ is used in
  the denominator in equation (\ref{eq:rat}).},  
\be
R(k)=\frac{P_{k,ab}}{P_{k,aa}}=C\sqrt{\frac{P_{k,bb}}{P_{k,aa}}}, 
\label{eq:rat}
\ee 
can be calculated.

Of course, in the real ICM, instead of pure isobaric, adiabatic or isothermal perturbations, we
 have a mixture of different types of perturbations\footnote{Possibly a continuum of
values of $\zeta_i$ in equation (\ref{eq:flux_rat}).}. Using equations (3) and (4), it is easy to
rewrite expected values of $C$ and $R$ for any given mix of these
three types of perturbations via the power spectra $S_{k,i}$ of the density fluctuation fields $(\delta n/n)_i$, assuming that these fields are uncorrelated.  Calculating $C$ and $R$ in terms of this decomposition, we find
\be
C(k)=\frac{\sum\limits_{i}S_{k,i}w_{a,i}w_{b,i}}{\sqrt{\sum\limits_i
     S_{k,i} w^2_{a,i}}\sqrt{\sum\limits_i S_{k,i}w^2_{b,i}}}=\frac{\sum\limits_{i}\alpha_i^2w_{a,i}w_{b,i}}{\sqrt{\sum\limits_i
     \alpha_i^2 w^2_{a,i}}\sqrt{\sum\limits_i \alpha_i^2 w^2_{b,i}}},
\label{eq:coh1}
\ee
and
\be
R(k)=\frac{\sum\limits_{i}S_{k,i}w_{a,i}w_{b,i}}{\sum\limits_{i}S_{k,i}w_{a,i}^2}=\frac{\sum\limits_{i}\alpha_i^2w_{a,i}w_{b,i}}{\sum\limits_{i}\alpha_i^2w_{a,i}^2},
\label{eq:rat1}
\ee
where $\alpha_i^2=S_{k,i}/\sum\limits_j S_{k,j}$ are normalized spectra.
Clearly, if one type of fluctuations dominates,
then $C=1$ and $R$ coincides with the values given in equation
(\ref{eq:flux_rat}) and shown in
Fig. \ref{fig:flux_nonlin}. Otherwise, $|C|<1$ and $R$ has the value
intermediate between these curves.

Since $\alpha_{\rm isob.}^2+\alpha_{\rm adiab.}^2+\alpha_{\rm isoth.}^2=1$, the maps of the expected values
$C$ and $R$ can be calculated as functions of two parameters $\alpha_{\rm isob.}$
and $\alpha_{\rm adiab.}$, setting $\alpha_{\rm  isoth.}=\sqrt{1-\alpha^2_{\rm isob.}-\alpha^2_{\rm adiab.}}$ for any
combination of the considered three types of perturbations. Then,
measuring $C$ and $R$ through the observed power and cross spectra of
emissivity fluctuations, equations (\ref{eq:coh}) and (\ref{eq:rat}),
and finding these values on $C$ and $R$ maps, the relative
contribution of each type of perturbations to the observed total variance of the
fluctuations at a given wavenumber $k$ can be obtained.

If the amplitude of fluctuations is large, $\sim$ few tens
per cent, $C$ and $R$ in adiabatic and isobaric cases
may differ from those shown in Fig. \ref{fig:flux_nonlin}. Our
simulations (see Appendix \ref{app:nonlin}) show that as long as the
amplitude of density fluctuations is $\lesssim 15$ per cent in gas with
temperature $> 3$ keV, the ratio is consistent with the values
in the limit of small-amplitude perturbations. 

\section{Data processing, images and power spectra}
\label{sec:processing}

We use public {\it Chandra} data of the Perseus Cluster with total
cleaned exposure $\approx 1.4$ Ms. We assume the redshift of the Perseus Cluster $z=0.01755$ and the angular diameter distance $71.5$ Mpc. 1 arcmin corresponds to a
physical scale $20.81$ kpc. The total Galactic HI column density is
$1.35\times10^{21}$ cm$^{-2}$ \citep{Dic90,Kal05}. The initial data reduction, the
details of image processing and the treatment of point sources are
described in detail in \citet[][Section 2]{Zhu15}. Both images, in
the $0.5-4$ keV and $4-8$ keV bands, are treated identically. Masking out
point sources and the central AGN, the best-fitting
spherically symmetric $\beta$ models of surface brightness are
obtained. Their parameters are $r_c=1.28\,(1.48)$ arcmin and $\beta=0.53
\,(0.49)$ in the soft (hard) band.  The images of Perseus divided by
the corresponding underlying $\beta-$model profiles in both bands are shown in
Fig. \ref{fig:images}. There are many structures produced by gas
perturbations of different nature. The large-scale structures whose nature
was identified through ``X-arithmetic'' analysis (Churazov et
al., in prep.) are marked with arrows.
        
The power and cross-spectra of surface brightness fluctuations in the X-ray images are
measured using the modified $\Delta$ variance method
\citep{Oss08,Are12}. Subtracting Poisson noise and correcting for the
suppression factor associated with $2D\to3D$ deprojection, which
depends only on the global geometry of the cluster, the 3D power
spectra of the volume emissivity fluctuations in both bands, $P_{k,aa}$
and $P_{k,bb}$, and the cross spectrum $P_{k,ab}$ are
calculated. Instead of power spectra, we will show characteristic
amplitude defined as $A_{k,ab}=\sqrt{4\pi P_{k,ab}k^3}$ (similarly
for $A_{k,aa}$ and $A_{k,bb}$) that is a proxy for the RMS of
emissivity fluctuations $\delta f/f$ at a given wavenumber $k=1/l$.
For the data set considered below, the values of $P_{k,ab}$ are always
positive.

Similar analysis was already applied to the analysis of
density fluctuations in the AWM7 Cluster \citep{San12}, the Coma Cluster
\citep{Chu12}, the Perseus Cluster \citep{Zhu15}, the Virgo Cluster
\citep{Are15} and the Centaurus Cluster \citep{Wal15}. Slightly
different approaches have been used to measure pressure fluctuations
in the Coma Cluster \citep{Sch04} and temperature fluctuations in a
sample of clusters \citep{Gu09}. The only difference here from our
previous analyzes is that here the $2D\to 3D$ deprojection is calculated at each pixel
instead of averaged over an area of interest. The
underlying $\beta$ models are slightly different in two bands since
the gas temperature changes with radius. Therefore, we calculate the
geometrical correction (the suppression factor) individually for each band. Uncertainties
associated with the choice of the underlying $\beta$ model and inhomogeneous
exposure coverage \citep[see details in ][Section 6]{Zhu15} are
estimated and taken into account for each fluctuation spectrum
presented here.

The analysis of X-ray surface brightness fluctuations directly
  measures fluctuations of the volume emissivity  convolved with {\it
    Chandra} response. In the soft band $a$, where the temperature
  dependence of X-ray emissivity is weak, $P_{k,aa}\approx 4 S_{k,aa}$, provided that $\delta n/n \ll 1$ \citep{Chu12}. In the hard band,
the temperature dependence is significant, and, therefore, the spectrum of the emissivity fluctuations
does not correspond to the spectrum of pure density fluctuations.

\section{Results}

\begin{figure*}
\includegraphics[trim=0 0 0 0,width=\textwidth]{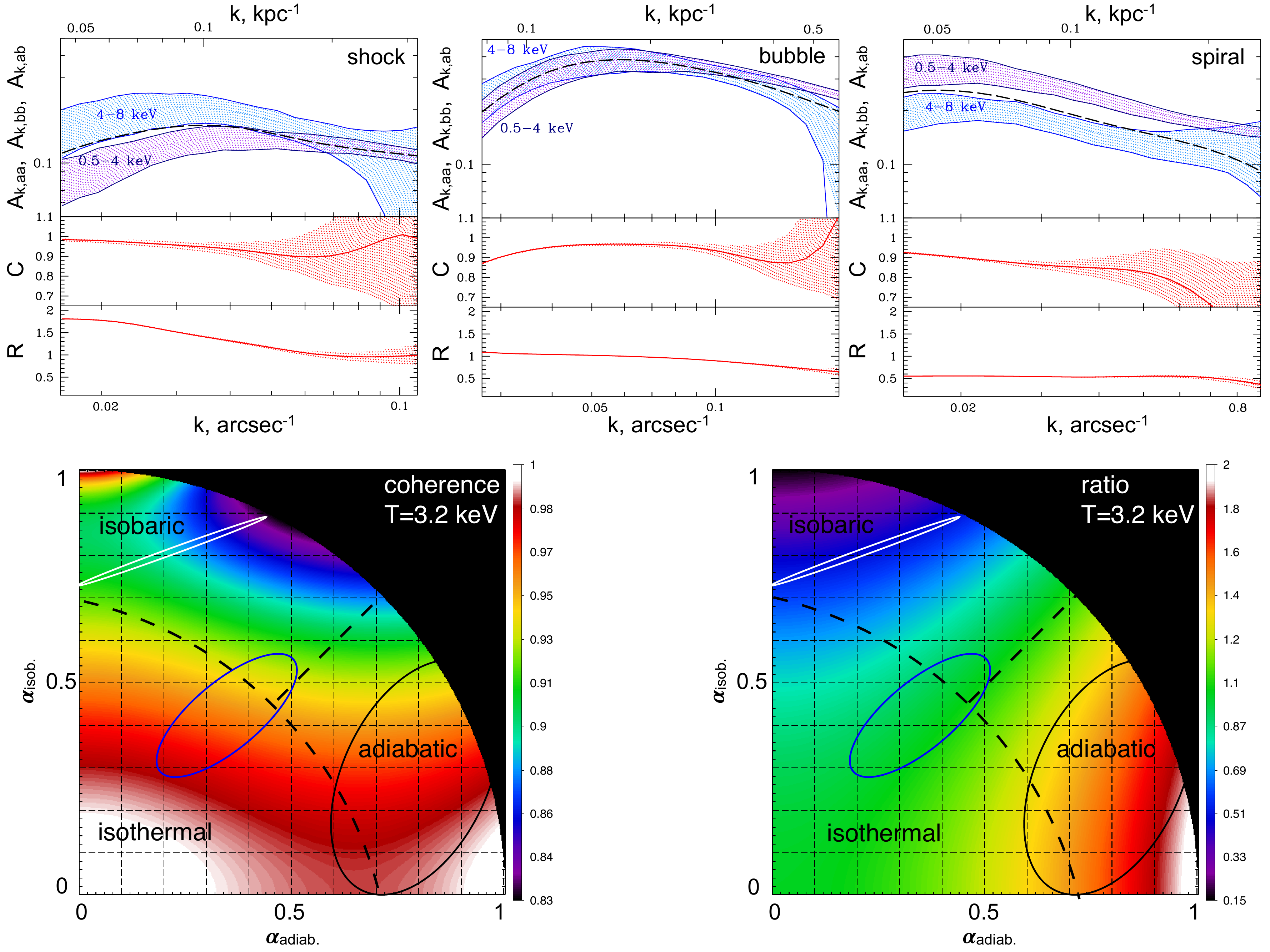}
\caption{Results of the cross-spectra analysis in regions with the
  shock (top left), bubble (top middle) and spiral (top right) (see
  Figure \ref{fig:images}). {\bf Top row:} amplitude of the volume emissivity
  fluctuations, in the soft, $A_{k,aa}$, (purple) and hard, $A_{k,bb}$, (blue) bands
  and the cross-amplitude, $A_{k,ab}$, (black dashed curve); coherence $C$ and
  ratio $R$ obtained from the observed spectra [equations
  (\ref{eq:coh}) and (\ref{eq:rat})]. Blue and purple dotted regions
  show $1\sigma$ statistical and stochastic uncertainties. For clarity,
  we do not plot the uncertainties on the cross-amplitudes. Our
  conservative estimates of statistical uncertainties on measured $C$
  and $R$ are shown with the dotted red regions. {\bf Bottom row:}
  maps of coherence $C$ (left) and ratio $R$ (right) for a
  mixture of isobaric, adiabatic and isothermal perturbations in the $3.2$
  keV gas. Color bars show the values of $C$ and $R$ [equations (\ref{eq:coh1}) and (\ref{eq:rat1})]. The dashed grid
  shows the values of $\alpha_i$, associated with each type of fluctuations (see Section \ref{sec:ampl}). X-axis:
  contribution of adiabatic fluctuations $\alpha_{\rm adiab.}$; Y-axis:
  contribution of isobaric fluctuations $\alpha_{\rm isob.}$. The contribution
  of isothermal fluctuations is
  $\alpha_{\rm isoth.}=\sqrt{1-\alpha_{\rm adiab.}^2-\alpha_{\rm isob.}^2}$. The maps are
  schematically divided into three regions where one of the types of perturbations is dominant in terms of total variance. Ellipses show the regions of
  $\alpha_{\rm isob.}$, $\alpha_{\rm adiab.}$ and $\alpha_{\rm isoth.}$ that correspond to the values
  of $R$ and $C$ taken from the figures in the top row. Black ellipse:
  the region with shock, blue ellipse: bubble, white ellipse: spiral. The size of each
  ellipse reflects the uncertainties associated with Poisson
  noise, the choice of the underlying model and the choice of the
  weighting scheme in calculating the power spectra. The locus of $C$
  and $R$ is in the adiabatic area if measured in the region
  with the shock in Perseus, in the isobaric area if obtained from the
  region with spiral structure and in the isothermal area when the region
  with the bubble is considered.
\label{fig:tests}
}
\end{figure*}

\begin{figure*}
\includegraphics[trim=100 0 100 0,angle=90,width=\textwidth]{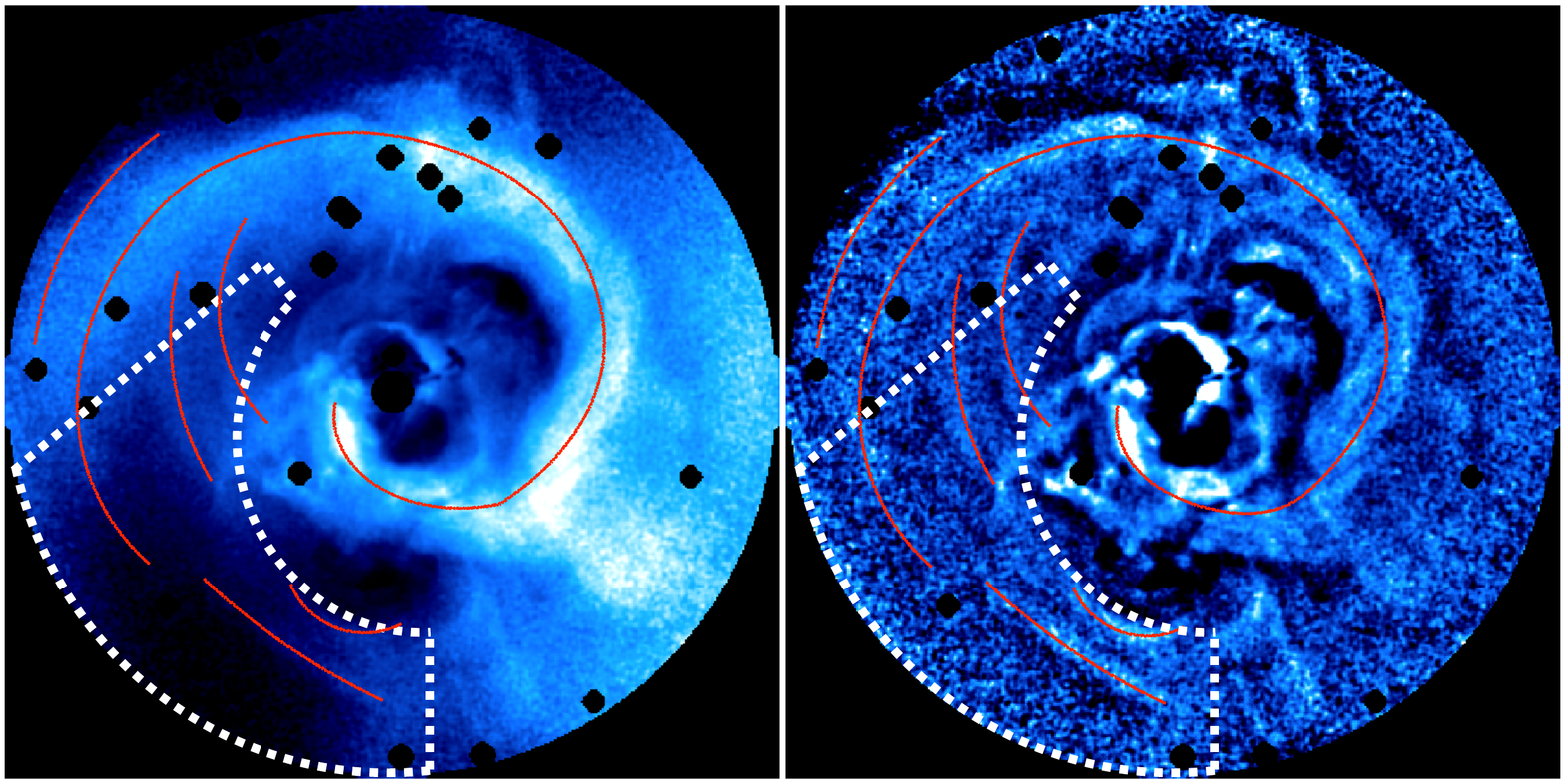}
\caption{Image of the central $r=3.5$ arcmin ($\approx 70$ kpc) region of the Perseus Cluster divided by the spherically-symmetric $\beta$ model of the surface brightness (left) and additionally filtered with the high-pass filter (right); see details in Section 6.1 in  \citet{Zhu14b}. Point sources and the central AGN are excised. The most prominent ripples are schematically highlighted with red curves. A selected region of the cluster with ripples and without obvious bubbles, central shock, filaments or prominent spiral structure, is shown with dotted boundaries. This region is used in Section \ref{sec:cen_reg} to test the properties of this subset of ripples. 
\label{fig:central_im}
}
\end{figure*}

Here we first test our analysis technique in selected regions of the Perseus Cluster that are dominated by prominent previously identified structures,
such as a shock, a bubble or a spiral (Section
\ref{sec:spec_cases}). Next, we apply the same analysis
to fluctuations in the central ($7\times 7$
arcmin) strongly perturbed by AGN activity, as well as in the region dominated by ripple-like structures
(Section \ref{sec:cen_reg}). For each of these cases, we
investigate the effects of systematic uncertainties, such as
inhomogeneous exposure coverage and the choice of the underlying model
(for details, see Section 6 of \citealt{Zhu15}). We use a uniform weighting
scheme  when calculating RMS of fluctuations present in filtered images, and a spherically symmetric $\beta$ model as the
underlying model. Our experiments show that various systematic
uncertainties do not change our conclusions, although some specific
numbers may differ for different weightings or underlying models.

We also tested the cross-spectra technique on simulated X-ray
  images containing shocks and sound waves (see Appendix
  \ref{sec:test_simul}). These tests show that our analysis is able to
  recover the nature of the dominant type of fluctuations even if the amplitude of the fluctuations is small, $\sim$ a few per cent.

\subsection{Special cases: shock, bubble, spiral}
\label{sec:spec_cases}
Residual images of three selected regions are shown in
Fig. \ref{fig:images}. All three regions are within the innermost
$30$ kpc, where the number of photon counts is large and the gas
temperature is $\approx 3-3.4$ keV. For a mean temperature of $3.2$
keV, we calculate the $C$ and $R$ maps for all possible combinations of
isothermal, adiabatic and isobaric perturbations
(Fig. \ref{fig:tests}, bottom panels). Amplitudes of the volume emissivity
fluctuations in the soft and hard bands and cross-amplitude as functions of wavenumber are shown in the top panels of Fig. \ref{fig:tests}.

In the region with a shock (spiral),
the amplitude in the hard band is larger (smaller) than in the soft
band over a broad range of scales. Emissivity fluctuations have
comparable amplitudes in both bands in the ``bubble'' region.

Fluctuations associated with the shock region have $C\simeq0.90-0.98$ and $R\simeq1.3-1.8$ on
scales between $8$ and $20$ kpc. The locus of these values on the $C$ and $R$ maps is shown
with the black ellipses. They lie within the area dominated by adiabatic perturbations.

Emissivity fluctuations in the region containing the bubble have $C\simeq0.92-0.98$ and $R\simeq0.90-1.05$ on scales between $4$ and $10$ kpc. The
locus of these values (blue ellipse in Fig. \ref{fig:tests}) reveals
a predominantly isothermal nature of these perturbations with small contamination of adiabatic and isobaric fluctuations,
associated with the shock around the bubble and displaced gas, respectively, imprints of which
remain in the selected region.

In the region with a spiral structure, we find $C\simeq0.8-0.9$ and
$R\simeq0.50-0.51$ on $10-20$ kpc scales. The locus of these values (white
ellipse in Fig. \ref{fig:tests}) confirms the isobaric nature of these
 fluctuations.

These experiments show that even if the
uncertainties are large, the cross-spectra method robustly identifies the
dominant type of fluctuations in the X-ray emissivity.

\subsection{Central, AGN-dominated region}
\label{sec:cen_reg}  

\begin{figure*}
\includegraphics[trim=60 0 60 0,width=\textwidth]{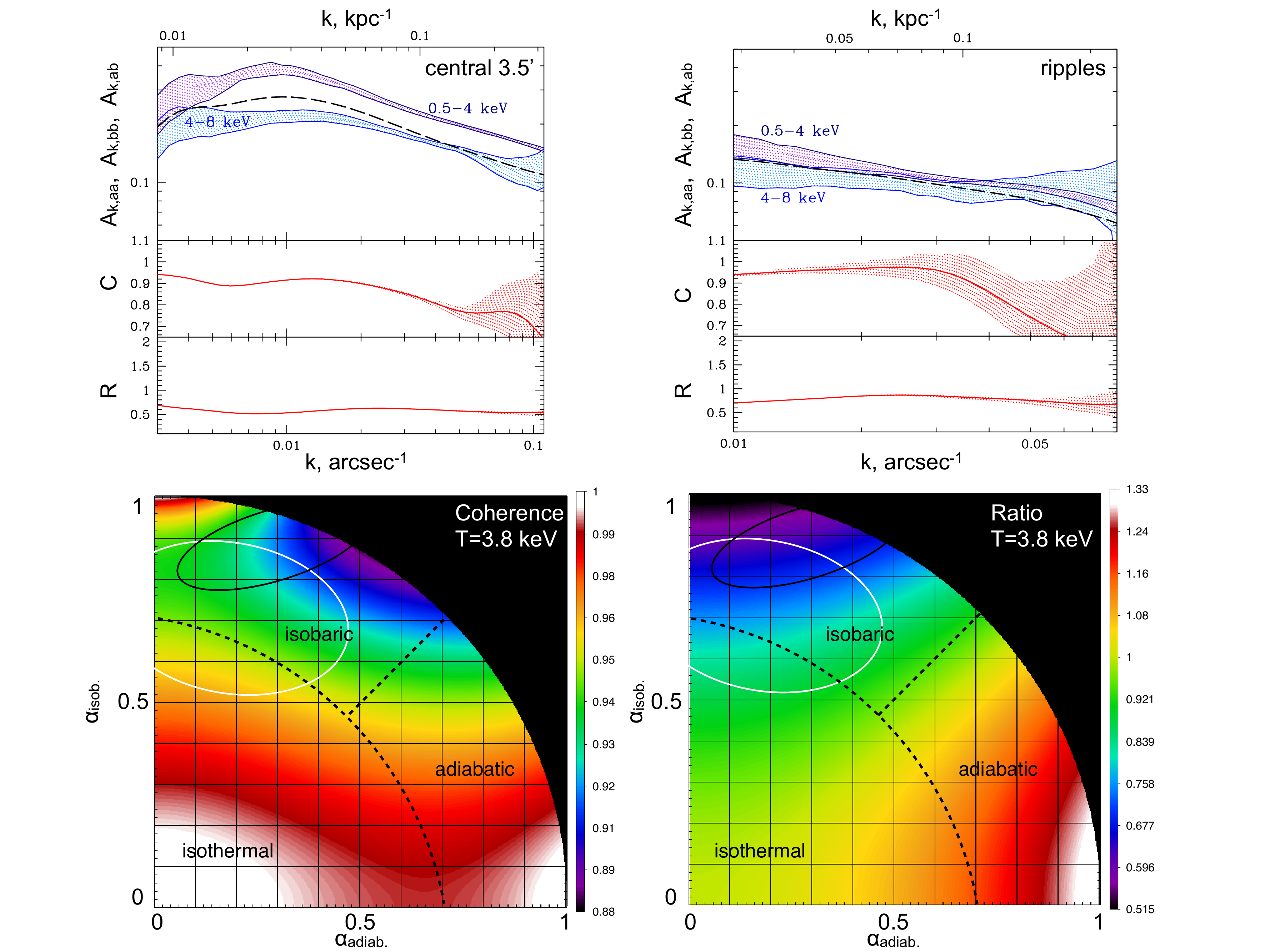}
\caption{Results of the cross-spectra analysis in the core of the
  Perseus Cluster shown in Fig. \ref{fig:central_im}. {\bf Left:}
  results from the entire central $r\sim 3.5$ arcmin ($\sim 70$ kpc)
  region. {\bf Right:} results from the selected dotted region
  (Fig. \ref{fig:central_im}). {\bf Top row:} amplitude of the volume
  emissivity fluctuations in the soft and hard bands,
  cross-amplitudes, measured coherence $C$ and ratio $R$. {\bf Bottom
    row:} the corresponding $C$ and $R$ maps. Notation
  and color coding are the same as in Fig. \ref{fig:tests}.  The
  ellipses show a range of parameters that lead to the observed values
  of $C$ and $R$. The sizes of the ellipses reflect statistical and
  systematic uncertainties. Black ellipse: the entire $r=3.5$ arcmin
  region, white ellipse: the selected dotted region
  (Fig. \ref{fig:central_im}). The fluctuations in the central region are predominantly isobaric.
\label{fig:central}
}
\end{figure*}

We now proceed with the analysis of the central $r=3.5$ arcmin
($\approx 70$ kpc) region, which appears to be disturbed by several
types of perturbations and is
dominated by a prominent spiral structure and bubbles
(Fig. \ref{fig:central_im}). Masking out point sources and the
central AGN, we measure the amplitude of the volume emissivity fluctuations in
both $0.5-4$ keV and $4-8$ keV bands, the coherence $C$
and the ratio $R$ shown in the top panels in Fig. \ref{fig:central}. Black ellipse on the $C- R$ map in
the bottom panel in Fig. \ref{fig:central} shows the locus of the measured values
$C\simeq0.80-0.93$ and $R\simeq0.50-0.55$ on scales $\sim 8-70$ kpc. 
The approximate center of the locus gives $\alpha_{\rm adiab.}\simeq 0.28$, $\alpha_{\rm isob.}\simeq 0.88$ and the corresponding $\alpha_{\rm isoth.}=\disp\sqrt{1-\alpha_{\rm adiab.}^2-\alpha_{\rm isob.}^2}\simeq 0.38$. Therefore, $\alpha_{\rm isob.}^2\approx 80$ per cent of the total variance can be attributed to
isobaric type of fluctuations, less than $\alpha_{\rm adiab.}^2\approx 8$ per cent of the variance to
adiabatic and $\alpha_{\rm isoth.}^2\approx 14$ per cent to isothermal perturbations. Exclusion of the innermost
bubbles and shocks from the central $r=3.5$ arcmin region tilts
the results towards even larger isobaric fraction.

The unsharp masking of the Perseus image shows approximate concentric
features, so-called ripples, that are narrow in the radial direction
and wide in the azimuthal direction \citep{Fab06, San07}. The width of
the ripples in the radial direction is roughly $5-15$ kpc (see,
e.g., Fig. \ref{fig:central_im}). Some of the
ripples are associated with the brightest and most clearly defined parts
of the spiral structure and have an isobaric nature. However, the
nature of those ripples that are not associated with the spiral structure is
not clear. Two possible scenarios for their origin have been discussed:
weak shocks and sound waves propagating through the gas \citep[see
  e.g.][and references therein]{San07,Gra08,Fab11} or stratified
turbulence \citep{Zhu14,Zhu15}. Aiming to understand the physical
origin of the ripples, we repeated the cross-spectra analysis of the
emissivity fluctuations in the region from which we excised the ripples
associated with prominent spiral, bubbles, shocks and filaments - see
the dotted region in Fig. \ref{fig:central_im}. The results, shown in Fig. \ref{fig:central} (white ellipse), reveal that the mean value of $\alpha_{\rm adiab.}\simeq 0.23$, $\alpha_{\rm isob.}\simeq 0.7$ and $\alpha_{\rm isoth.}\simeq 0.67$. Therefore, $\alpha_{\rm isob.}^2\approx 50$ per cent of the total variance is associated with isobaric, $\alpha_{\rm adiab.}^2\approx 5$ with adiabatic and $\alpha_{\rm isoth.}^2\approx 45$ per cent with isothermal perturbations on scales $\sim 12-30$ kpc. In other words, the variance presumably induced by the
slow motions of the gas in the cluster atmosphere and bubbles constitutes the
largest fraction of the total energy in this region on scales larger
than 12 kpc. This conclusion does not exclude the presence of ripples
associated with shocks and sound waves. However, energetically such
perturbations appear to be subdominant. An interesting question would be
to understand the nature of fluctuations on smaller scales, less than
12 kpc. Currently the high level of Poisson noise in the hard band
limits our ability to measure the amplitude of emissivity fluctuations
on such small scales. For the same reason, it is difficult to perform
the analysis for individual ripples. To address these
questions, at least twice longer {\it Chandra} observations would be
needed.

Similar analysis applied to a region outside the inner part of Perseus, the annulus at $3.5-6$ arcmin, shows that isobaric fluctuations
are again the dominant contributor to the total variance on scales $\sim 20-50$ kpc. In this annulus, the photon statistics in the hard band are too low to constrain the nature of perturbations on smaller scales.

\section{Discussion}

\subsection{Energetics of AGN-driven perturbations and their dissipation timescale}
Measured variance of the volume emissivity fluctuations can be translated into the total energy $E_{\rm pert}$ associated with
AGN-induced perturbations in the gas, which we write as $E_{\rm
  pert}=E_{\rm b}+E_{\rm sw}+E_{\rm gw}$, where $E_{\rm b}$ is the
energy in the fluctuations due to bubbles of relativistic plasma
(isothermal perturbations), $E_{\rm sw}$ is the energy associated with
sound waves and shocks (adiabatic perturbations) and $E_{\rm gw}$ is
the energy in fluctuations induced by gravity waves, turbulence or any
other slow displacements of the gas (isobaric perturbations). All other contributions are ignored for simplicity. It is
convenient to compare these energies to the thermal energy
of the gas $E_{\rm th}=PV/(\gamma-1)=3PV/2$, where $P$ is its pressure
and $V$ is the volume of the region under consideration. Clearly, for each type of perturbation $E/E_{\rm th}\sim \langle(\delta n/n)^2\rangle\approx \frac{1}{4} \langle(\delta f/f)^2\rangle$ if $\delta f/f$ is measured in soft (density) band and the perturbation amplitude is small. Let us estimate the proportionality coefficient between $E/E_{\rm th}$ and $\delta n/n$ in each case.

The total energy of a
bubble is a sum of the internal energy and work done by the expanding
bubble on the cluster gas, i.e., $E_{\rm
  b}=\gamma_{\rm b}PV_{\rm b}/(\gamma_{\rm b}-1)$, where {\bf $\gamma_{\rm b}=4/3$} is the
adiabatic index of the hot relativistic gas inside the bubble and $V_{\rm b}$ is
the volume of the bubbles \citep[e.g.,][]{Chu01}. The ratio $E_{\rm
  b}/E_{\rm th}=\gamma_{\rm b}(\gamma-1)X_{\rm b}/(\gamma_{\rm b}-1)$, where $X_{\rm b}$ is the
fraction of the volume occupied by the bubbles.  Assuming that the
bubbles are completely devoid of thermal gas, they will
correspond to density fluctuations $\delta n/n=-1$,
while fluctuations in the rest of the gas are small. Hence, if the bubbles occupy a small fraction of the volume, the total variance of the gas density fluctuations in any given volume becomes
$\langle\left(\delta n/ n\right)^2_{\rm
  b}\rangle=X_{\rm b}$, and, therefore \citep[see also][]{Are15},
\be  \frac{E_{\rm b}}{E_{\rm
    th}}=\disp\frac{\gamma_{\rm b}(\gamma-1)}{\gamma_{\rm b}-1}\bigl\langle\left(\frac{\delta n}{n}\right)^2_{\rm
  b}\bigl\rangle.
\label{eq:b}
\ee

 The total energy in sound waves is
\be
E_{\rm sw}=\int \left(\frac{\rho v^2}{2}+\frac{c_s^2\delta\rho^2}{2\rho}\right)dV,
\label{eq:ewave}
\ee 
where $\rho=\mu m_p n$ is the gas density, $\mu=0.61$ is the mean
particle weight, $m_p$ is the proton mass, $v$ is the velocity of the gas in the wave and $c_s$ is the
sound speed of the gas \citep{Lan59}. For the propagating linear plane wave $\delta \rho/\rho=\delta n/n=v/c_s$, and, therefore, the ratio of the energy in sound waves to the thermal energy is
 \be \frac{E_{\rm sw}}{E_{\rm
    th}}=\gamma(\gamma-1)\bigl\langle\left(\frac{\delta n}{n}\right)^2_{\rm
  sw}\bigl\rangle.
\label{eq:sw}
\ee

The energy of gravity waves is
\be
E_{\rm gw}=\int \left(\frac{\rho v^2}{2}+\frac{H_p}{H_s}\frac{c_s^2\delta\rho^2}{2\rho}\right)dV,
\label{eq:gwave}
\ee
where $H_p$ and $H_s$ are pressure and entropy scale heights respectively, the ratio of which is $H_p/H_s=1/(\gamma-1)$ in isothermal cluster.
 In the case of propagating gravity waves the density perturbations $\langle\left(\delta n/n\right)^2_{\rm
  gw}\rangle\propto(\gamma-1)\langle\left(v/c_s\right)^2\rangle$ on scales larger than the Ozmidov scale \citep[see, e.g.,][]{Zhu14b}. Therefore, accounting for equation (\ref{eq:gwave}), the ratio of the energy in gravity waves to the thermal energy becomes
\be
\frac{E_{\rm gw}}{E_{\rm th}}=\gamma \bigl\langle\left(\frac{\delta n}{n}\right)^2_{\rm
  gw}\bigl\rangle.
\label{eq:gw}
\ee

Since the coefficient between energy ratios and the variance of
density fluctuations is a factor of a few in all cases [equations (\ref{eq:b}), (\ref{eq:sw}) and (\ref{eq:gw})] and the measured total variance of the fluctuations
attributed to isobaric type of perturbations constitutes $\approx 80$ per cent of the
total variance in the inner $r=3.5$ arcmin in Perseus (Section
\ref{sec:cen_reg}), we will use equation (\ref{eq:gw}) to convert the variance of density fluctuations to the total energy in perturbations.

We obtain the variance $\langle\left(\delta n/n\right)^2\rangle$
by integrating the measured 3D power spectrum of the volume emissivity fluctuations $P_{k,aa}$ in the soft (``density'') band
\be
\bigl\langle\left(\frac{\delta n}{n}\right)^2\bigl\rangle=\frac{1}{2}\int\limits_{k_{\rm
    min}}^{k_{\rm max}}P_{k,aa}4\pi k^2 dk.
\label{eq:int}
\ee
$P_{k,aa}$ converted to the amplitude $A_{k,aa}$ is shown in the left
panel in Fig. \ref{fig:central}. The integration over the entire range
of scales that we probe with our measurements gives
$\langle\left(\delta n/n\right)^2\rangle\sim 0.08$, which implies that the nonthermal energy is $\sim 13$ per cent of the thermal energy. This estimate should be considered as a lower limit.

\citet{Zhu14} showed that there is an approximate balance between
local radiative cooling rate and turbulent dissipation rate in the
core of the Perseus Cluster. Therefore, one can write
$E_{\rm pert}/t_{\rm diss}\approx E_{\rm
    th}/t_{\rm cool}$, where $t_{\rm diss}$ is the dissipation
timescale needed to convert the non-thermal energy into heat in
order to maintain the balance of cooling and turbulent energy
dissipation and
\be
t_{\rm cool}=\frac{3}{2}\frac{(n_e+n_i)k_{\rm B}T}{2n_e n_i \Lambda (T)}
\ee
is the cooling time ($n_e$ and $n_i$ are the number
densities of electrons and ions respectively, $k_{\rm B}$ if the Boltzmann constant). In the inner $r=3.5$
arcmin in Perseus, the cooling time varies between $0.45$ and $3$ Gyr
\citep[e.g.,][]{Zhu14}. Therefore, the dissipation time is given by
{\rm $t_{\rm
  diss}\sim t_{\rm cool}E_{\rm pert}/E_{\rm
    th}\approx 0.06 - 0.4$} Gyr. This is consistent with the
dissipation timescale in the core of the Virgo Cluster, which was
estimated using the same technique by \citet{Are15}.

\begin{figure*}
\includegraphics[trim=0 280 0 30, width=\textwidth]{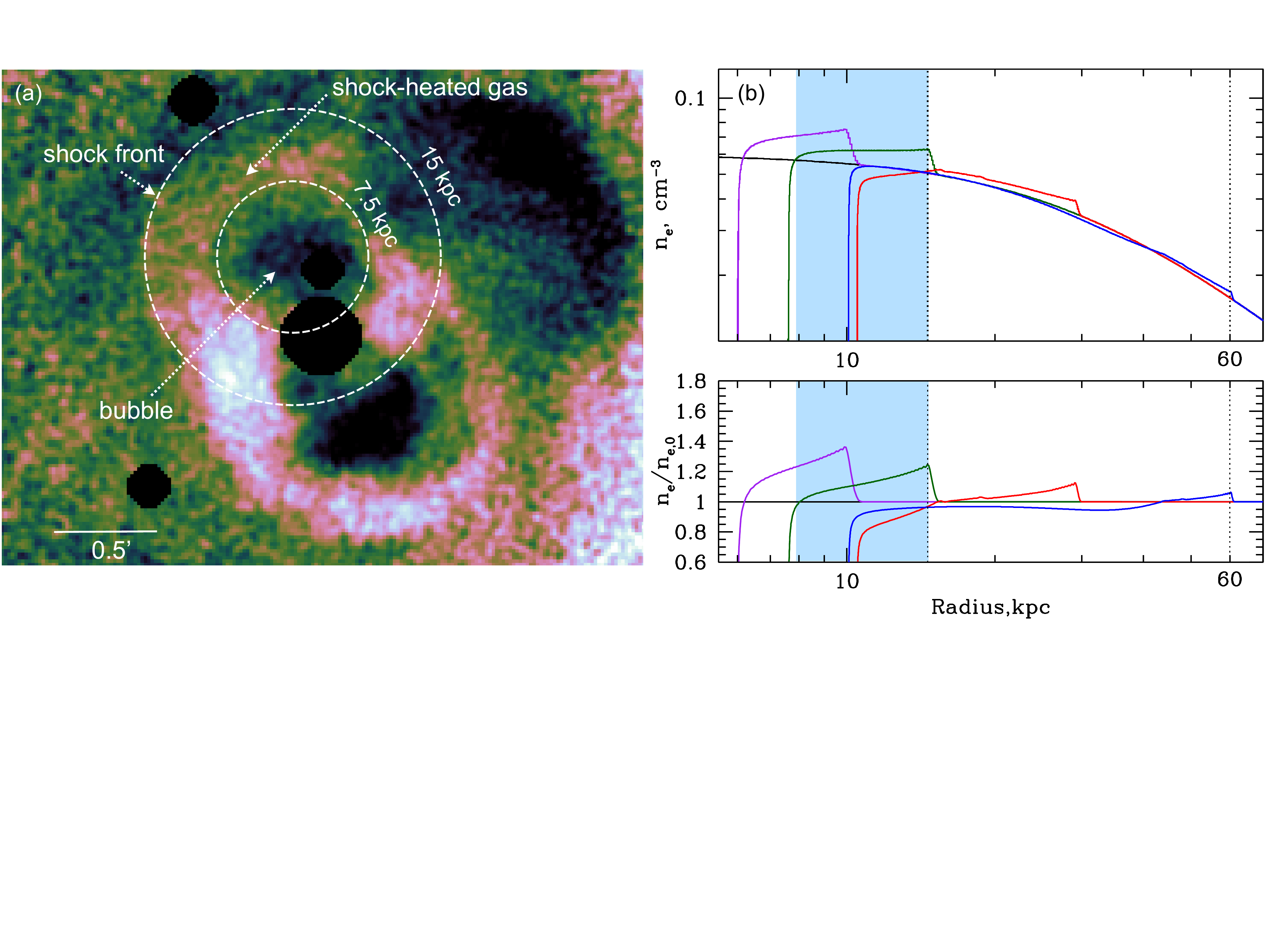}
\caption{Features produced by the AGN outburst in the innermost region in the Perseus Cluster. {\bf (a):} the image of the innermost region of the Perseus
  Cluster ($\sim 60\times 60$ kpc) in the 4-8 keV band divided by the
  best-fitting $\beta$ model. Point sources and the central AGN are
  excised. Dashed circles indicate one of the inner bubbles with the
  size $r \approx 7.5$ kpc and a shock front around it at $r \approx 15$ kpc
  from the bubble center.  {\bf (b):} Snapshots of simulated density
  profiles perturbed by a spherical shock, which propagates through
  the gas in the Perseus Cluster. The shock is driven by an outburst
  with power $3\cdot 10^{44}$~\ergss. Black curve shows the initial
  (unperturbed) density profile, while the colour curves show the
  profiles at different times after the energy injection starts. The
  dotted vertical lines indicate the radii of $15$ and $60$
  kpc. The former corresponds to the position of the shock around the
  inner bubble in Perseus, while the latter shows the approximate
  location of the distinct ripples. For the shock at $\approx 15$ kpc, the
  size of the region with the shock-heated gas (shell) is indicated
  with the shaded blue region. The ratio of the perturbed density to
  the initial one is shown in the bottom panel.
\label{fig:fake_shock_den}
}
\end{figure*}

\subsection{The partition of energy from the central AGN outburst}
\label{sec:shock_simul}
\begin{figure}
\includegraphics[trim=60 100 210 40,width=0.5\textwidth]{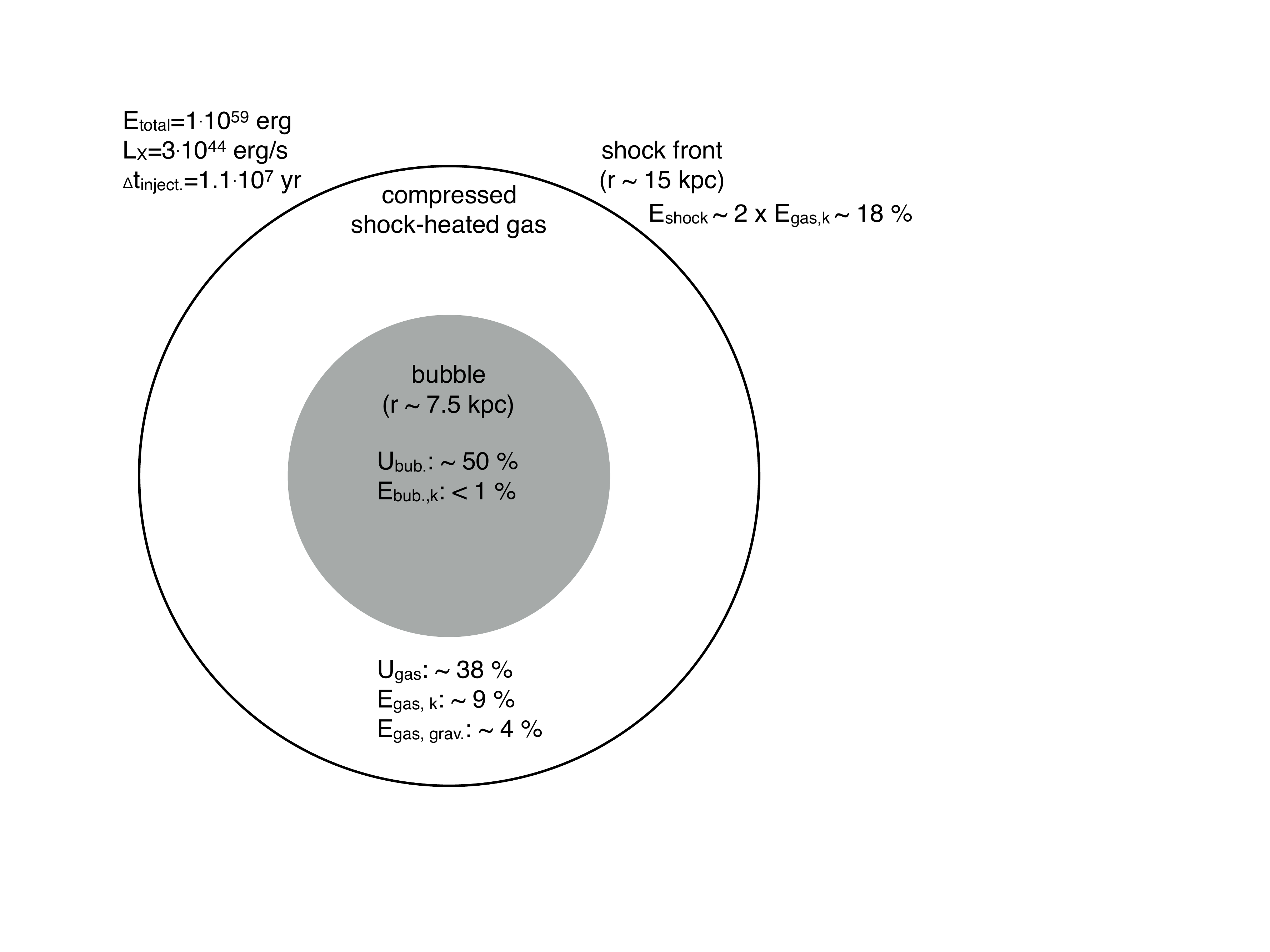}
\caption{Sketch of the AGN-injected energy partition in the inner
  $15$ kpc in the Perseus Cluster $\Delta t_{\rm inject.}=1.1\cdot 10^7$ yr after the
  injection started. The results are based on 1D simulations of a
  spherical shock propagating through the cluster atmosphere. The
  energy is injected at a constant rate $L_X=3\cdot 10^{44}$ erg/s. The total energy released by the AGN at this particular moment (when shock is at $15$ kpc from the bubble center) is $E_{\rm total}\approx 10^{59}$ erg. The
  input parameters are shown in the top left corner. About $50$ and
  $38$ per cent of the outflow energy is channelled to the excess
  thermal energies in bubble and in shock-heated gas respectively. See
  Section \ref{sec:shock_simul} for details.
\label{fig:energy_distr}
}
\end{figure}

\begin{table*}
\centering
\caption{Summary of the observed features in the innermost region in
  the Perseus Cluster (Fig. \ref{fig:fake_shock_den}) and the same
  parameters obtained from 1D simulations of the spherical shock
  propagating through the Perseus atmosphere, see Section
  \ref{sec:shock_simul} for details.}
\begin{tabular}{@{}lll@{}}
\hline
&Observed &  Simulated\\ 
\hline
Injected jet power per bubble, $L_X$ &$\approx 5\cdot 10^{44}$ erg/s$^{(1)}$& $ 3\cdot 10^{44}$ erg/s \\
Position of the shock front relative to the bubble center& $\approx 15$ kpc$^{(2)}$ & $ 15 $ kpc\\
Amplitude of the density jump at the shock front & $ 1.25-1.27$$^{(3)}$ & $ 1.25$\\
Ratio of the shell (shock-heated gas) size to the bubble size, $r_{\rm shell}/r_{\rm b}$ & $\approx 2^{(2)}$ & $1.9$\\
\hline
\multicolumn{2}{l} {\footnotesize(1) \citet{Boe93,Hei98,Chu00,Fab00}} \\
\multicolumn{2}{l}{\footnotesize(2) estimated from the images, see Fig. \ref{fig:fake_shock_den}}\\
\multicolumn{2}{l}{\footnotesize (3) \citet{Gra08b}}\\
\hline
\label{tab:param}
\end{tabular}
\end{table*}
We showed above that isobaric perturbations are energetically dominant in the inner region of the Perseus Cluster, where gas physics is strongly affected by the central AGN. If these isobaric perturbations are associated with slow gas displacements induced by buoyantly rising and expanding bubbles of relativistic plasma, then the energy from the AGN outburst should be partitioned so that the largest fraction of it goes to the bubbles rather than to shocks or the compressed, shock-heated gas around the bubbles. The partition of the energy depends on the duration of the outburst and on the total outburst energy. \citet{For15} discuss two extreme scenarios: a
short-duration outburst, which produces small bubbles surrounded by hot, low-density gas and strong shocks, which carry out most of the AGN energy, and  a long-duration outburst, which produces weaker shocks and larger bubbles, which store most of the AGN energy surrounded by the cooler gas.  Using 1D numerical simulations
of a spherical shock propagating into the gas, \citet{For15} showed
that the observational properties of the most prominent shock in the
center of the M87/Virgo Cluster are best described with a long-outburst
model with total energy outburst $\sim 5\cdot 10^{57}$ erg lasting
$\sim 2$ Myr. In this model, $\approx 50$ per cent of the injected energy
goes into the thermal energy of the bubble and $\approx 20 $ per cent is
carried away by the shock. Admittedly, the modeling is done under
strong assumptions about the source geometry, initial density
and temperature profiles, (non-) constant energy injection rate,
etc. However, \citet{For15} estimated that the expected uncertainties are
only a factor of a few. Below we perform similar analysis for the
central shock and bubble in the Perseus Cluster, aiming to understand whether indeed most of the AGN energy is injected into the bubbles. This energy can potentially be later transferred to the gas in the form of isobaric perturbations. 

There are several features revealed by deep {\it Chandra} observations
of the innermost region in Perseus produced by the outburst
(Fig. \ref{fig:fake_shock_den}a): inner bubbles with size $r\approx 7.5$
kpc, distinct shock around one of the bubbles at $r\approx 15$ kpc from
the bubble center and a shell of shock-heated gas between the
bubble and the front. Modeling of the observed properties of X-ray cavities in the central
$\sim 25$ kpc in the Perseus Cluster induced by the relativistic
particles of the jet implies a time-averaged nuclear power of the order of $10^{45}$~\ergss
\citep[e.g., ][]{Hei98,Chu00,Fab00}. The deprojection analysis of the gas density shows a density jump at
the shock front is $1.25-1.27$ \citep{Gra08b}. Using 1D hydro simulations of a spherical shock propagating through the Perseus atmosphere, we will try to reproduce these observed features and check the energy partition that results in these simulations.

As initial, unperturbed radial profiles of the gas density and temperature in Perseus, we used the deprojected ones from \citet[][]{Zhu13b}. Of course, the initial profiles that might have existed more than 10 Myrs ago are quite uncertain. However, if the supermassive black hole is able to maintain a quasi-equilibrium between heating and radiative cooling and the cluster atmosphere has not undergone major changes recently, then we can hope that the present gas distribution is not far from the typical conditions at the time of the outburst. The power of the AGN is chosen so that the size of the density jump at the
shock front and the ratio of the shell size to the bubble size $r_{\rm
  shell}/r_{\rm b}$ approximately match the observed
values. Namely, we assume a constant energy injection rate into the
bubble $L_X=3\cdot 10^{44}$ erg/s (roughly consistent with earlier
estimates $\approx 5\cdot 10^{44}$ erg/s per bubble, see, e.g., \citet{Hei98}) lasting for $2
\cdot 10^{7}$ yrs. The energy injection is quenched when the bubble
expansion becomes subsonic and the bubble size is $r\approx 10$
kpc. Subsequently, the shock propagates passively through the cluster
atmosphere. As an illustrative example, we
show the snapshots of the density radial profiles at different time
steps and the ratio of the perturbed density to the initial one in
Fig. \ref{fig:fake_shock_den}b. At $\Delta t_{\rm inject.}=1.1\cdot 10^7$ yr after the injection started, the parameters of the simulated features are the closest to the observed ones; see Table \ref{tab:param}. Accounting for uncertainties in measurements of the observed features, our conservative estimate of the lower limit on the duration of energy injection is $7$ Myrs.  Note that $5$ per cent uncertainties in the density jump and bubble size translate into $\approx 20$ uncertainties in the injected power.

Our simulations provide information about the redistribution of the
injected outburst energy at each time step. When the shock is at $15$ kpc from the bubble center, the total energy released by the AGN
is $\approx 10^{59}$ erg. This energy is partitioned so that it gives an
excess of thermal energy within the bubble of $\approx 5\cdot
10^{58}$ erg and in the shock-heated gas of $\approx 4\cdot 10^{58}$ erg
(Fig. \ref{fig:energy_distr}). After the shock propagation, the total
gravitational energy increases by $\approx 4 \cdot 10^{57}$ erg and the
total kinetic energy is $\approx 9.5\cdot 10^{57}$ erg in the entire
$r= 15$ kpc region. This kinetic energy is mostly associated with
the kinetic energy of the heated and compressed gas. Note that $\approx
50$ per cent of the outburst energy goes into the excess thermal energy (compared to initial thermal energy) of the bubble and $\approx 38$ per cent, to the excess thermal energy of the compressed and shock-heated
gas. The energy in the bubble is larger than in the shell by a factor of
$1.3$ (the factor is $2.6$ if one assumes that the gas inside
  the bubble is relativistic with the adibatic index $\gamma=4/3$). Such energy partition supports the
slow-outburst scenario of the energy injection and bubble
inflation\footnote{Our experiments showed that for a given total injected energy $\sim 10^{59}$ erg, the shortening of the outburst duration leads to a stronger heating of the shock-heated gas. Timescales shorter than $10^6$ yrs correspond to the limiting case of the short-outburst scenario.}. Finally, the energy carried by the shock is twice
larger than the kinetic energy of the heated gas, namely $\approx
1.9\cdot10^{58}$ erg, or $\approx 18$ per cent of the injected energy
from the AGN (additional energy is associated with the thermal energy
of the compressed gas). Note that although the energetics and parameters of the jet, shock
and bubble are different from those in the Virgo Cluster \citep{For15}, the way the
energy from the central AGN is injected into the gas is the same in
both clusters.

Our modeling implies that indeed the largest fraction of the outburst energy
is stored in bubbles and a smaller fraction is transported by a
shock. This result relies on the assumption that our 1D approximation
and initial conditions provide a reasonable description of the cavity
dynamics and the energy partition can be estimated directly from the
model. A somewhat different conclusion was reached by \citet{Gra08},
who used an extrapolation of the pressure profile from the unshocked
region to smaller radii to estimate the excess energy. Their approach
yields a significantly larger fraction of energy in the shock,
suggesting that shock/sound waves dominate transport of energy to
the ICM. Our results instead provide support to the bubble-mediated
AGN feedback model
\citep{Chu00,McN00,Nul07,For07,Fab12,Bir12,Hla12}. Namely, most of the
energy is first stored as the enthalpy of the bubbles. As the bubbles rise buoyantly they transfer their energy to the ICM.
As we show here,
isobaric perturbations dominate the total variance of fluctuations,
suggesting that the energy stored within the bubbles could be
channeled to the ICM through gravity waves and gas motions  induced
during the bubble expansion and buoyant rise. The energy of the gas bulk motions
eventually dissipates, providing enough heat to offset radiative
cooling \citep{Zhu14b}. The data of future high-energy-resolution
missions starting from ASTRO-H will be instrumental in differentiating
between the two energy-transporting scenarios mentioned above.

We also looked at the properties of a sound wave produced by the
central AGN propagating through the Perseus atmosphere after
 the energy release is quenching. Simulations show that the amplitude of a passively propagating
sound wave leads to density jumps of $\approx 5 $ per cent at a distance
of $60$ kpc, which are consistent with those associated with the ripples in
Perseus.  At this moment, the energy is partitioned so that the contribution of the kinetic energy of these waves to
the total injected energy is small, less than $8$ per cent.

\section{Conclusions}
In this work, we investigated the nature of AGN-driven perturbations in
the core of the Perseus Cluster and determined the energetically-dominant
type of perturbations based on the statistical analysis of the X-ray emissivity fluctuations
in soft- and hard-band images obtained from deep {\it Chandra}
observations. We also examined how energy injected by the AGN is partitioned between different feedback channels using 1D hydro simulations of a shock wave propagating through the Perseus Cluster atmosphere. Our main results are summarized below.
\begin{enumerate}
\item{The analysis of cross-spectra is able to recover the dominant
  type of fluctuations responsible for the fluctuations in the X-ray emissivity
  even if their amplitude is small, of order of a few per cent. This is
  confirmed by various tests that we have carried out using both simulated
  and observed data.}
\item Most of the emissivity fluctuations induced by the central AGN in the inner $\approx 140\times140$ kpc
  region have an isobaric nature ($\approx 80$ per cent of the total
  variance) on scales $\sim 8-70$ kpc, i.e., are consistent with entropy variations that can result from slow displacements of the gas. 
\item In the selected region where there are distinct ripples least
  contaminated by other structures, about $50$ per cent of the
  total variance of the fluctuations is associated with isobaric
  perturbations on scales $\sim 12-30$ kpc, supporting the stratified turbulence 
  nature of the ripples \citep{Zhu14b}. This conclusion does not exclude the presence
  of ripples associated with shocks or sound waves, however, they are
  energetically subdominant ($\approx 5$ per cent). In order to
  investigate the nature of individual ripples as well as to probe
  smaller scales using the cross-spectra analysis, at least twice longer {\it Chandra}
  observations are required. Currently,
  low photon statistics in the hard band limit the analysis.
\item Using the fact that the total variance of perturbations is proportional to the
  energy in perturbations, we estimated that the non-thermal energy is $\sim 13$ per cent of the thermal energy in the inner
  $r=70$ kpc region. 
\item We also estimated the time needed to dissipate  the energy from the observed
  fluctuations
  into heat, to balance radiative cooling. This time is $\approx 6\cdot 10^7 - 4\cdot
  10^8$ yr.
\item { Simulating the observed properties of the inner shock, bubble and shock-heated gas around it in Perseus, we found that $\approx 50$ per cent of the AGN-injected energy goes into the excess thermal energy of the bubble, $\approx 40$ per
  cent goes into the compression and heating of the gas between the
  bubble and the shock front and $\approx 20$ of the injected energy is carried away by the shock. These results are consistent with
  the bubble-mediated AGN feedback model and support our main results from the cross-spectrum analysis, i.e., the energetically-dominant role of slow gas motions that could be induced by the bubbles in the AGN feedback.}
\item{ Our simulations of the propagating shock show that the contribution of sound waves at distance of $60$ kpc from the cluster center (ripples) to the total excess energy is small, less than $8$ per cent. However, a typical amplitude of the simulated ripples of a few per cent is consistent with observations.}
\end{enumerate}

The small number of photons, especially in the hard band, limits the
analysis presented in this work. The full potential of this method will be
unveiled with future X-ray observations that have large effective area
(a few tens of times larger than {\it Chandra}), such as {\it
  Athena}\footnote{http://www.the-athena-x-ray-observatory.eu/} and
{\it X-ray
  Surveyor}\footnote{http://cxc.cfa.harvard.edu/cdo/xray\_surveyor/}. Having
arcsecond angular resolution with a few eV spectral resolution, both
provided by {\it X-ray Surveyor}, will be especially helpful for probing the
physics of fluctuations.

\section{Acknowledgements}
Support for this work was provided by the NASA through Chandra
award number AR4-15013X issued by the Chandra X-ray Observatory Center, which is operated by the Smithsonian Astrophysical Observatory for and on behalf of the NASA under contract
NAS8-03060. IZ thanks A.C. Fabian and J. Sanders for important comments and discussions. EC and AAS are grateful to the Wolfgang Pauli Institute, Vienna, for its hospitality. PA acknowledges support from FONDECYT grant 1140304 and Centro de Astrofisica de Valparaiso. We are grateful to Hans-Thomas Janka for the provision of a one-dimensional Lagrangian hydrodynamics code, initially used for supernovae explosion calculations.

\appendix
\section{Choice of energy bands}
\label {enbands}
Depending on the choice of bands, the flux ratios [Fig. \ref{fig:flux_nonlin}, equation (\ref{eq:flux_rat})] in pure adiabatic and isobaric cases shift closer to or further away from each other. The harder the hard band and/or the larger the gap between the soft and hard bands, the larger the difference between the adiabatic and isobaric ratios. Therefore, the energy bands of the X-ray images used in the analysis are chosen based on two simple criteria: (i) the larger the difference between the adiabatic and isobaric curves in Fig. \ref{fig:flux_nonlin} the more powerful the diagnostic of fluctuations of different types, i.e. the difference
\be
\Delta w=\frac{w_{\rm b,adiab}}{w_{\rm a,adiab.}}-\frac{w_{\rm
    b,isob.}}{w_{\rm a,isob.}}
\ee
is large; (ii) bands should be broad enough so that the uncertainties in the cross spectrum due to noise are small. In the limit of a noise-dominated data the ability to detect the difference between adiabatic and isobaric fluctuations depends on the ratio $\Delta w/\sigma_a\sigma_b$, where  $\sigma_{a(b)}\propto 1/ \Lambda_{a(b)}^{1/2}$ are the Poisson errors in the images, divided by the smooth model.  Therefore, those bands that give the maximal value of the ratio
\be
\frac{\Delta w}{\sigma_a\sigma_b}= \left[\frac{w_{\rm b,adiab}}{w_{\rm
      a,adiab.}}-\frac{w_{\rm b,isob.}}{w_{\rm a,isob.}}\right] \left (\Lambda_a \Lambda_b \right )^{1/2}
\label{eq:crit}
\ee
 should be chosen. Both criteria showed that the $0.5-4$ keV and $4-8$ keV combination is the optimal choice. Experiments with other, similarly good bands, gave results consistent with those presented in the main text.

\section{$C$ and $R$ in the case of large-amplitude fluctuations}
\label{app:nonlin}
\begin{figure}
\includegraphics[trim=0 160 0 100,width=0.49\textwidth]{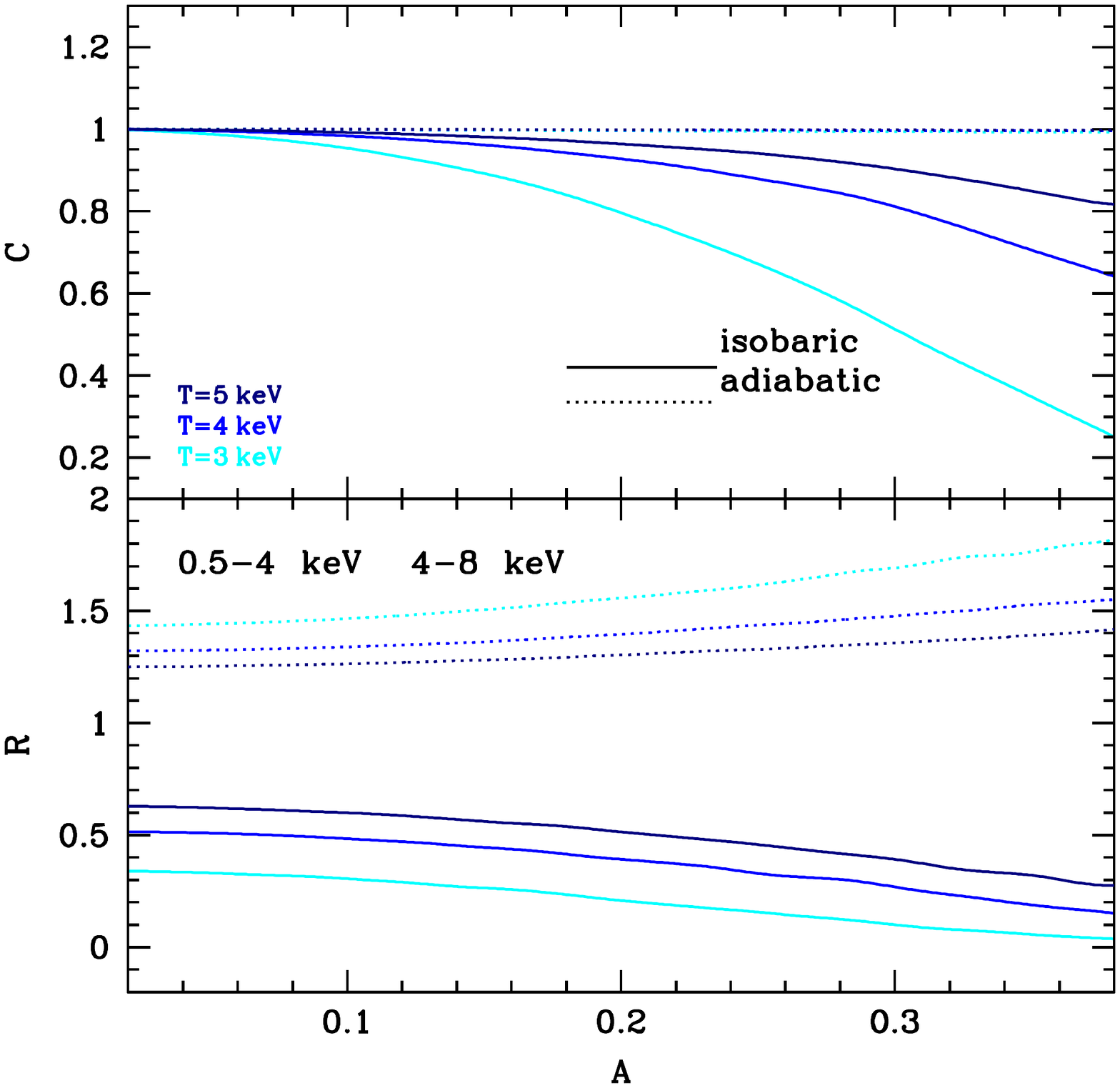}
\caption{Coherence $C$ and ratio $R$ of emissivity fluctuations in two
  energy bands, $0.5-4$ keV and $4-8$ keV as a function of a parameter
  $A$, which characterizes the amplitude of log-normal density fluctuations modeled by equation (\ref{eq:den_simul}), for three different
  gas temperatures in the cases of pure isobaric or adiabatic
  fluctuations. Note that in the adiabatic case, $C$ does
  not vary much with the amplitude of fluctuations (as expected) or with
  gas temperature, in contrast to the isobaric case. For $A\lesssim
  0.15$ both $C$ and $R$ only weakly depend on $A$ if the gas is hotter than $3$ keV. 
\label{fig:nonlin}
}
\end{figure}

Maps of coherence $C$ and ratio $R$ (Figs. \ref{fig:tests}, \ref{fig:central})
are calculated under the assumption of small amplitude of
fluctuations. In order to check how robust these predictions for
larger amplitudes are, we carried out simple simulations. Assuming
that density fluctuations along the line of sight have a
log-normal distribution as suggested by numerical simulations
\citep{Kaw07,Zhu13}, we generated density fluctuation patterns as
\be
\disp\frac{n+\delta n}{n}=e^{A\xi},
\label{eq:den_simul}
\ee
where $A$ is a parameter characterizing the amplitude of fluctuations
and $\xi$ is a random, normally distributed variable with a zero mean and standard deviation 1. Corresponding temperature
fluctuations are $(T+\delta T)/T=[(n+
\delta n)/n]^{\zeta_i-1}$, where $\zeta_i=5/3, 0$ and $1$ for adiabatic, isobaric and isothermal
fluctuations, respectively. Assuming that there are many fluctuations along
the line of sight, we obtain emissivity perturbations $\delta f/f$ in the soft and hard
bands and measure $C$ and $R$ shown in Fig. \ref{fig:nonlin}. Both $C$ and $R$ are
consistent with the values in the limit of small-amplitude
perturbations [equation (\ref{eq:flux_rat}), Figs. \ref{fig:flux_nonlin}, \ref{fig:tests} and \ref{fig:central}] as long as the value of $A$ in the gas with $T>3$ keV is $\lesssim 15$
per cent. Also, $C$ in
the cases of pure isobaric and adiabatic fluctuations is almost the
same, while $R$ is different in all three cases. Therefore, $R$ is a
better diagnostic for differentiation between different types of
perturbation in the gas with Perseus-like temperatures than $C$.

\section{Cross-spectra analysis of simulated images with shock and sound waves}
\label{sec:test_simul}
\begin{figure}
\includegraphics[trim=20 350 10 100,width=0.5\textwidth]{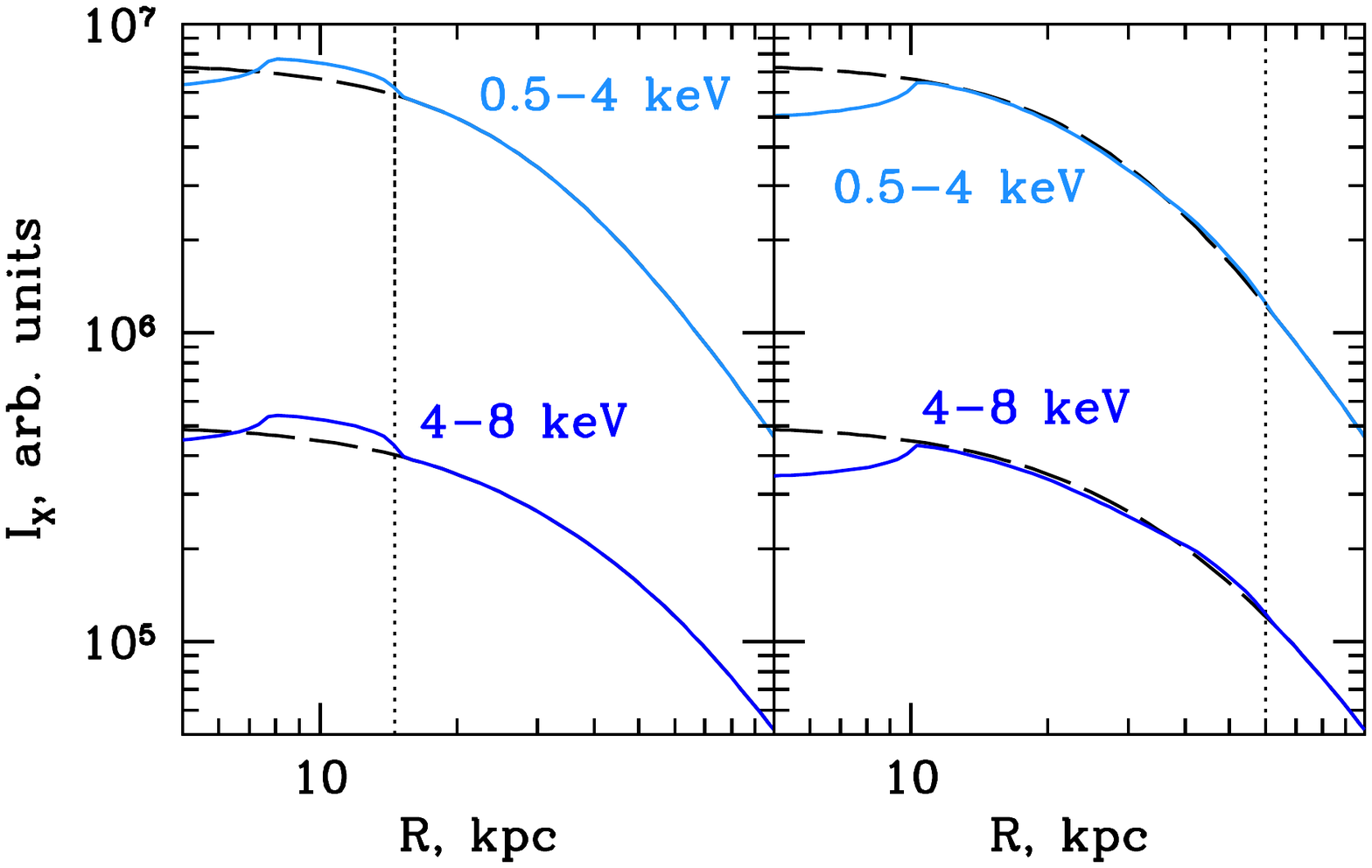}
\caption{Simulated radial profiles of the X-ray surface brightness $I_X$ in
  the $0.5-4$ keV and $4-8$ keV bands, calculated from the density and
  temperature radial profiles modified by the propagating spherical
  shock in the Perseus Cluster (see
  Fig. \ref{fig:fake_shock_den}b). The unperturbed profiles are shown
  with black dashed curves. {\bf Left:} shock at $15$ kpc from the
  bubble center. {\bf Right:} shock at $60$ kpc.
\label{fig:fake_sb}
}
\end{figure}
\begin{figure*}
\includegraphics[trim=0 260 0 120,width=1.\textwidth]{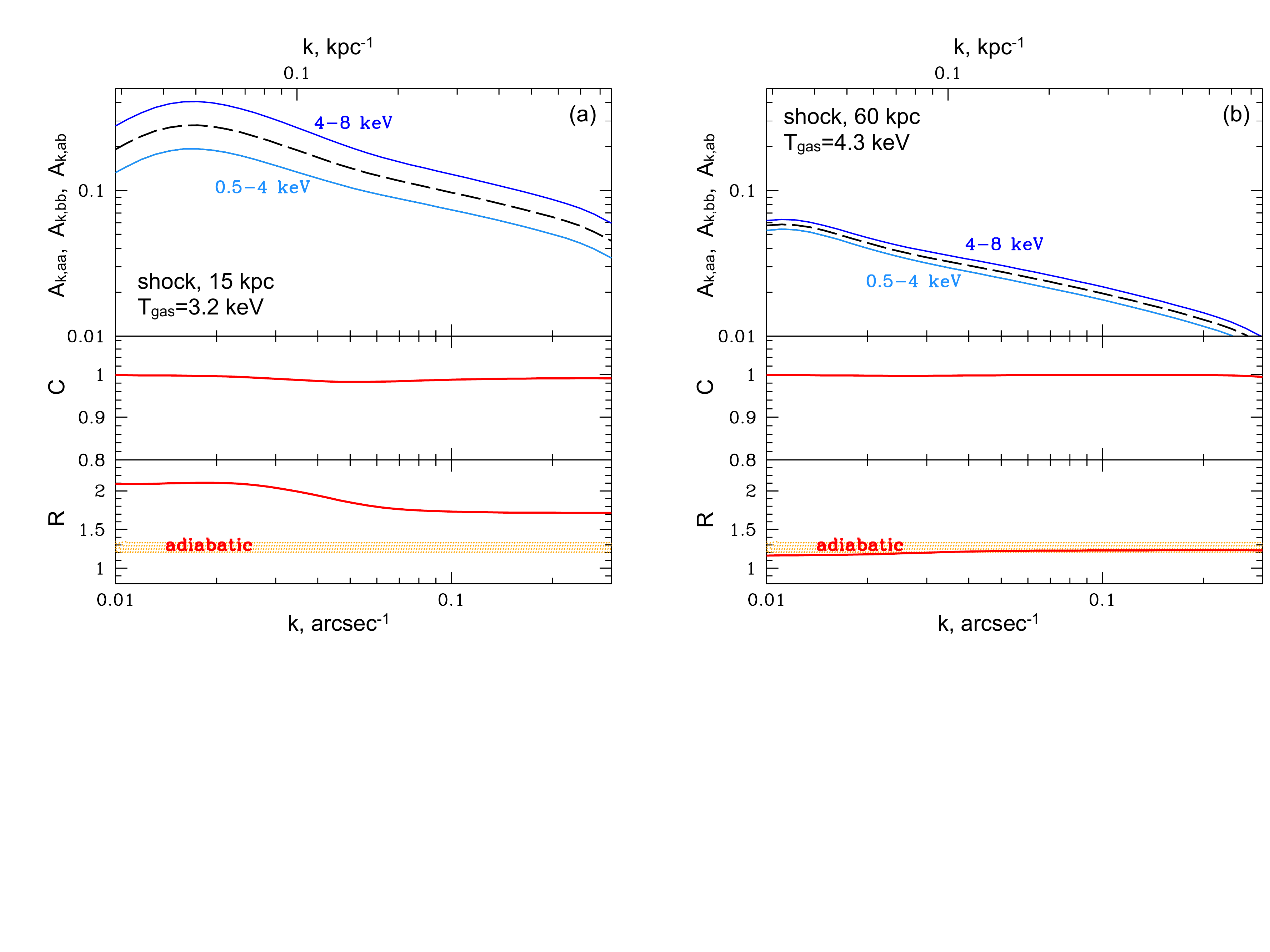}
\caption{Cross-spectra analysis of simulated images with a shock/sound
  wave (see Figs. \ref{fig:fake_shock_den}b, \ref{fig:fake_sb}). The
  amplitude of emissivity fluctuations in both bands and the cross-amplitude are
  shown in the top panels. Correlation coefficient $C$ and ratio $R$
  are shown in the middle and bottom panels, respectively. Expected $R$ (Fig. \ref{fig:flux_nonlin}) for pure adiabatic fluctuations is shown with
  orange regions. {\bf (a):} shock at $15$ kpc. The analysis is
  done in the annulus around the shock with the width $0.8$
  arcmin. {\bf (b):} sound wave at $60$ kpc. The width of the
  annulus used for the analysis is $1.6$ arcmin. The cross-spectra
  analysis captures the adiabatic nature of the perturbation in both cases,
  even if the amplitude of the emissivity fluctuations is small (a few
  per cent).
\label{fig:fake_cross_sp}
}
\end{figure*}
An interesting question is whether the cross-spectra method is able
to determine correctly the dominant type of perturbations
if their amplitude is small, viz., a few per cent. Using simulated
density and temperature radial profiles shown in
Fig. \ref{fig:fake_shock_den}b, we obtained the X-ray images in soft
and hard bands when the shock was at $15$ kpc and 60
kpc (see Section \ref{sec:shock_simul}). Fig. \ref{fig:fake_sb} shows the radial profiles of the X-ray
surface brightness in both cases. Note a clear discontinuity in the
surface brightness when the shock is at $15$ kpc and a barely
noticeable kink when the shock is at $60$ kpc.

We fit the profiles with $\beta$ models, obtained the residual images of
the surface brightness fluctuations and repeated the cross-spectra
analysis by measuring the amplitude of the volume emissivity fluctuations in both
bands, the cross-amplitude, the correlation $C$ and the ratio $R$. The results are
shown in Fig. \ref{fig:fake_cross_sp}.

First, we analyzed fluctuations in the annulus with a width of $0.8$
arcmin around the shock at 15 kpc. The annulus width roughly matches
the size of the region with the shock in the observed images that we
considered earlier (see Fig. \ref{fig:images}). The
amplitude of emissivity fluctuations is $\approx 26$ per cent and $\approx
14$ per cent (Fig. \ref{fig:fake_cross_sp}) at $\sim 10$ kpc scales in the soft and hard bands,
respectively. These values are close to those obtained from
observations in the extraction region of similar size. If wider annulus is used, the amplitudes derived from simulations go down compared to observations because in the latter, other perturbations besides the shock front contribute to the surface-brightness fluctuations. In contrast, the
measured coherence and
ratio are very stable. The measured coherence $C \approx 1$
(Fig. \ref{fig:fake_cross_sp}) confirms the presence of only one
type of perturbations. The ratio $R\sim 1.9$ at 10 kpc scale is slightly higher than
the predicted value $1.4 - 1.6$ (Fig. \ref{fig:flux_nonlin}). This deviation can be due to the
temperature variations in the chosen region and the relatively large
amplitude of emissivity fluctuations. Also the choice of the
underlying model may affect the results, especially on large
scales. Nevertheless, we can confidently exclude perturbations of an isothermal ($R=1$)
or isobaric ($R<0.4$) nature.

For the sound wave at 60 kpc, we consider a broader annulus of
width $1.6$ arcmin, which roughly matches the size of the dotted
region in Fig. \ref{fig:central_im}. The measured amplitudes of the
emissivity fluctuations are $\sim 4.5$ and $\sim 6$ per
cent on scales of $\sim 20$ kpc in the soft and hard bands, respectively (Fig. \ref{fig:fake_cross_sp}), which are lower than the observed value,
$\sim 12$ per cent, in Perseus (Fig. \ref{fig:central}). The measured
amplitude strongly depends on the size of the region analyzed. However,
both $C\approx 1$ and $R\approx 1.2$ robustly confirm the adiabatic nature of
the perturbations. This test shows that the cross-spectra analysis
is powerful enough to determine the dominant type of perturbations
even if the amplitude of emissivity fluctuations is only a few per cent.

\label{lastpage}  

\begin{thebibliography}{99}
\bibitem[\protect\citeauthoryear{Anders 
\& Grevesse}{1989}]{And89} Anders E., Grevesse N., 1989, GeCoA, 53, 197 
\bibitem[\protect\citeauthoryear{Ar{\'e}valo et 
al.}{2015}]{Are15} Ar{\'e}valo P., Churazov E., Zhuravleva 
I., Forman W.~R., Jones C., 2015, arXiv:1508.00013 
\bibitem[\protect\citeauthoryear{Ar{\'e}valo et 
al.}{2012}]{Are12} Ar{\'e}valo P., Churazov E., Zhuravleva 
I., Hern{\'a}ndez-Monteagudo C., Revnivtsev M., 2012, MNRAS, 426, 1793 

\bibitem[\protect\citeauthoryear{B{\^i}rzan et 
al.}{2012}]{Bir12} B{\^i}rzan L., Rafferty D.~A., Nulsen 
P.~E.~J., McNamara B.~R., R{\"o}ttgering H.~J.~A., Wise M.~W., Mittal R., 
2012, MNRAS, 427, 3468 
\bibitem[\protect\citeauthoryear{Boehringer et 
al.}{1993}]{Boe93} Boehringer H., Voges W., Fabian A.~C., 
Edge A.~C., Neumann D.~M., 1993, MNRAS, 264, L25 

\bibitem[\protect\citeauthoryear{Churazov et 
al.}{2001}]{Chu01} Churazov E., Br{\"u}ggen M., Kaiser C.~R., 
B{\"o}hringer H., Forman W., 2001, ApJ, 554, 261 
\bibitem[\protect\citeauthoryear{Churazov et 
al.}{2000}]{Chu00} Churazov E., Forman W., Jones C., B{\"o}hringer H., 2000, A\&A, 356, 788 
\bibitem[\protect\citeauthoryear{Churazov et 
al.}{2002}]{Chu02} Churazov E., Sunyaev R., Forman W., 
B{\"o}hringer H., 2002, MNRAS, 332, 729 
\bibitem[\protect\citeauthoryear{Churazov et 
al.}{2012}]{Chu12} Churazov E., et al., 2012, MNRAS, 421, 
1123

\bibitem[\protect\citeauthoryear{Dickey 
\& Lockman}{1990}]{Dic90} Dickey J.~M., Lockman F.~J., 1990, ARA\&A, 28, 215 

\bibitem[\protect\citeauthoryear{Fabian et al.}{2003}]{Fab03} 
Fabian A.~C., Sanders J.~S., Allen S.~W., Crawford C.~S., Iwasawa K., 
Johnstone R.~M., Schmidt R.~W., Taylor G.~B., 2003, MNRAS, 344, L43
\bibitem[\protect\citeauthoryear{Fabian}{2012}]{Fab12} Fabian A.~C., 2012, ARA\&A, 50, 455 
\bibitem[\protect\citeauthoryear{Fabian et al.}{2000}]{Fab00} 
Fabian A.~C., et al., 2000, MNRAS, 318, L65 
\bibitem[\protect\citeauthoryear{Fabian et al.}{2011}]{Fab11} 
Fabian A.~C., et al., 2011, MNRAS, 418, 2154 
\bibitem[\protect\citeauthoryear{Fabian et al.}{2006}]{Fab06} 
Fabian A.~C., Sanders J.~S., Taylor G.~B., Allen S.~W., Crawford C.~S., 
Johnstone R.~M., Iwasawa K., 2006, MNRAS, 366, 417 

\bibitem[\protect\citeauthoryear{Forman et al.}{2007}]{For07} 
Forman W., et al., 2007, ApJ, 665, 1057
\bibitem[\protect\citeauthoryear{Forman et al.}{2015}]{For15} 
Forman W. et al., 2015, submitted

\bibitem[\protect\citeauthoryear{Gu et al.}{2009}]{Gu09} Gu 
L., et al., 2009, ApJ, 700, 1161 


\bibitem[\protect\citeauthoryear{Graham, Fabian, 
\& Sanders}{2008}]{Gra08} Graham J., Fabian A.~C., Sanders J.~S., 2008, MNRAS, 391, 1749 
\bibitem[\protect\citeauthoryear{Graham, Fabian, 
\& Sanders}{2008}]{Gra08b} Graham J., Fabian A.~C., Sanders J.~S., 2008, MNRAS, 386, 278 


\bibitem[\protect\citeauthoryear{Heinz, Reynolds, 
\& Begelman}{1998}]{Hei98} Heinz S., Reynolds C.~S., Begelman M.~C., 1998, ApJ, 501, 126 
\bibitem[\protect\citeauthoryear{Hillel 
\& Soker}{2014}]{Hil14} Hillel S., Soker N., 2014, MNRAS, 445, 4161 
\bibitem[\protect\citeauthoryear{Hlavacek-Larrondo et 
al.}{2012}]{Hla12} Hlavacek-Larrondo J., Fabian A.~C., Edge 
A.~C., Ebeling H., Sanders J.~S., Hogan M.~T., Taylor G.~B., 2012, MNRAS, 
421, 1360 



\bibitem[\protect\citeauthoryear{Kalberla et 
al.}{2005}]{Kal05} Kalberla P.~M.~W., Burton W.~B., Hartmann D., Arnal E.~M., Bajaja E., Morras R., P{\"o}ppel W.~G.~L., 2005, A\&A, 440, 775 
\bibitem[\protect\citeauthoryear{Kawahara et 
al.}{2007}]{Kaw07} Kawahara H., Suto Y., Kitayama T., Sasaki 
S., Shimizu M., Rasia E., Dolag K., 2007, ApJ, 659, 257 

\bibitem[\protect\citeauthoryear{Landau 
\& Lifshitz}{1959}]{Lan59} Landau L.~D., Lifshitz E.~M., 1959, Course of Theoretical Physics. Pergamon Press, Oxford  

\bibitem[\protect\citeauthoryear{McNamara et 
al.}{2000}]{McN00} McNamara B.~R., et al., 2000, ApJ, 534, 
L135 
\bibitem[\protect\citeauthoryear{McNamara 
\& Nulsen}{2007}]{McN07} McNamara B.~R., Nulsen P.~E.~J., 2007, ARA\&A, 45, 117 

\bibitem[\protect\citeauthoryear{Nulsen et al.}{2007}]{Nul07} 
Nulsen P., McNamara B.~R., David L.~P., Wise M.~W., Leahy J.~P., 2007, AAS, 
39, 148 

\bibitem[\protect\citeauthoryear{Omma et al.}{2004}]{Omm04} 
Omma H., Binney J., Bryan G., Slyz A., 2004, MNRAS, 348, 1105 
\bibitem[\protect\citeauthoryear{Ossenkopf, Krips, 
\& Stutzki}{2008}]{Oss08} Ossenkopf V., Krips M., Stutzki J., 2008, A\&A, 485, 917 

\bibitem[\protect\citeauthoryear{Randall et 
al.}{2015}]{Ran15} Randall S.~W., et al., 2015, ApJ, 805, 112 
\bibitem[\protect\citeauthoryear{Reynolds, Balbus, 
\& Schekochihin}{2015}]{Rey15} Reynolds C.~S., Balbus S.~A., Schekochihin A.~A., 2015, ApJ, 815, 41


\bibitem[\protect\citeauthoryear{Sanders 
\& Fabian}{2007}]{San07} Sanders J.~S., Fabian A.~C., 2007, MNRAS, 381, 1381 
\bibitem[\protect\citeauthoryear{Sanders 
\& Fabian}{2012}]{San12} Sanders J.~S., Fabian A.~C., 2012, MNRAS, 421, 726

\bibitem[\protect\citeauthoryear{Schuecker et 
al.}{2004}]{Sch04} Schuecker P., Finoguenov A., Miniati F., B{\"o}hringer H., Briel U.~G., 2004, A\&A, 426, 387 
\bibitem[\protect\citeauthoryear{Soker, Hillel, 
\& Sternberg}{2015}]{Sok15} Soker N., Hillel S., Sternberg A., 2015, arXiv:1504.07754 

\bibitem[\protect\citeauthoryear{Walker, Sanders, 
\& Fabian}{2015}]{Wal15} Walker S.~A., Sanders J.~S., Fabian A.~C., 2015, MNRAS, 453, 3699

\bibitem[\protect\citeauthoryear{Zhuravleva et 
al.}{2015}]{Zhu15} Zhuravleva I., et al., 2015, MNRAS, 450, 
4184 
\bibitem[\protect\citeauthoryear{Zhuravleva et 
al.}{2014}]{Zhu14} Zhuravleva I., et al., 2014, Nature, 515, 
85 
\bibitem[\protect\citeauthoryear{Zhuravleva et 
al.}{2014}]{Zhu14b} Zhuravleva I., et al., 2014, ApJ, 788, L13 
\bibitem[\protect\citeauthoryear{Zhuravleva et 
al.}{2013}]{Zhu13} Zhuravleva I., Churazov E., Kravtsov A., 
Lau E.~T., Nagai D., Sunyaev R., 2013, MNRAS, 428, 3274 
\bibitem[\protect\citeauthoryear{Zhuravleva et 
al.}{2013}]{Zhu13b} Zhuravleva I., et al., 2013, MNRAS, 435, 
3111 

\end{thebibliography}
\end{document}